\documentstyle[twocolumn,aps,prb,epsfig]{revtex}

\title{
Planar cyclotron motion in unidirectional superlattices \\
defined by strong magnetic and electric fields:\\
Traces of classical orbits in the energy spectrum
}
\author{ 
S. D. M. Zwerschke, A. Manolescu$^{\dagger}$, and R. R. Gerhardts}
\address{
Max-Planck-Institut f\"ur Festk\"orperforschung, Heisenbergstra\ss e 1, 
D-70569 Stuttgart, Germany \\
$^{\dagger}$Institutul Na\c{t}ional de Fizica Materialelor, C.P. MG-7 
Bucure\c{s}ti-M\u{a}gurele, Rom\^ania}

\begin{document}

\maketitle

\begin{abstract}
  We compare the quantum and the classical description of the two-dimensional
  motion of electrons subjected to a perpendicular magnetic field and a
  one-dimensional lateral superlattice defined by spatially periodic magnetic
  and electric fields of large amplitudes.  We explain in detail the complicated
  energy spectra, consisting of superimposed branches of strong and of weak
  dispersion, by the correspondence between the respective eigenstates and the
  ``channeled'' and ``drifting'' orbits of the classical description.
\end{abstract}

\section{Introduction}
In the last decade there has been a constant interest in the transport
properties of the periodically modulated two dimensional electron gas (2DEG).
In particular, in the presence of a lateral modulation of a one-dimensional
character the resistivity may be strongly anisotropic, which essentially
reflects the anisotropy of the electronic states.  Two types of modulations can
be achieved in the experimental devices: electrostatic potential modulations
\cite{Weiss89:179,Gerhardts89:1173,Winkler89:1177,Zhang90:12850} and, more
recently, magnetic field modulations.
\cite{Carmona95:3009,Ye95:3013,Izawa95:706,Ye97:5444} Weak modulations of both
types lead already to pronounced magnetoresistance effects in the presence of an
average magnetic field $B_0$ applied perpendicular to the 2DEG. These effects
occur at low and intermediate $B_0$ values, well below the magnetic quantum
regime where Shubnikov-de Haas oscillations appear.  At very small values of
$B_0$ a pronounced positive magnetoresistance is observed, followed at
intermediate $B_0$ values by the ``Weiss oscillations'' due to commensurability
effects. Both effects are adequately understood within a classical transport
calculation based on Boltzmann's equation, and can be traced back to the
predominance of different types of classical trajectories.
\cite{Beenakker89:2020,Beton90:9229,Menne98:1707}

The positive magnetoresistance is understood as caused by ``channeled orbits''
which exist if the modulation is sufficiently strong or, equivalently, the
average magnetic field is sufficiently small. For electric modulation they occur
near the minima of the modulation potential (``open'' orbits
\cite{Beton90:9229}), and for magnetic modulation near the lines of vanishing
total magnetic field (``snake'' orbits \cite{Mueller92:385}). They are always
confined within a single period of the modulation, which we choose in $x$
direction.  They are wavy trajectories allowing for fast motion of electrons
with velocities within small angles around the direction of translational
invariance ($y$ direction).  These channeled orbits occur in addition to the
``drifting orbits'', which are self-intersecting trajectories with loops (along
each of which the direction of the velocity changes by $2\pi$), so that usually
a low drift velocity in $y$ direction results. For sufficiently small $B_0$,
drifting orbits may extend over many periods of the modulation. At sufficiently
large $B_0$ (sufficiently small modulation amplitudes) only the drifting orbits
survive.  The ``Weiss oscillations'' manifest a commensurability effect
depending on the ratio of the extent of drifting orbits (at the Fermi energy)
and the modulation period. With increasing modulation strength, the positive
magnetoresistance becomes more pronounced and extends to larger $B_0$ values,
suppressing progressively the low-$B_0$ Weiss oscillations.\cite{Beton90:9229}
This effect is well understood within the classical calculation
\cite{Menne98:1707}, if both types of trajectories are adequately included, and
it has recently also been obtained by a quantum calculation for a strong
modulation \cite{Shi96:12990}.

A qualitatively new type of magnetoresistance effect has recently been observed
by Ye {\em et al.}\cite{Ye98:330} on samples with an extremely strong magnetic
modulation. Samples with a surface array of ferromagnetic micro-strips were
measured in tilted magnetic fields, so that the applied magnetic field had a
large component parallel to the surface, producing a large magnetization of the
ferromagnetic strips, while only the small perpendicular component determined
the average magnetic field $B_0$ in the 2DEG. In this way a huge positive
magnetoresistance with superimposed Shubnikov-de Haas like oscillations was
obtained at low values of the average magnetic field, at which no magnetic
quantum effects should be expected for weak modulation.\cite{Ye98:330} It rather
seems that the quantum oscillations are induced by the large-amplitude periodic
magnetic modulation field.  Such conditions require a quantum transport theory
and, as a first step, the understanding of the quantum electronic states of a
2DEG with a strong magnetic modulation. This is the motivation of the present
work.

Channeled and drifting quantum states in linearly varying magnetic field are
already discussed by other authors.  \cite{Mueller92:385,Hofstetter96:4676} The
Schr\"odinger equation for periodic magnetic fields alternating in sign, has
been solved previously, but only for the case when the average field is zero
\cite{Ibrahim95:17321}.  In the present paper we shall study the quantum
electronic states in strong periodic magnetic fields with a non-vanishing
average, and compare it with the case of a strong electric modulation. In both
situations, rather complicated energy spectra are obtained, with striking
qualitative similarities and quantitative differences. For the case of strong
electric modulation, such a complicated energy spectrum has recently been
published \cite{Shi96:12990}, but without an attempt of an explanation.  We will
demonstrate in this paper that a close comparison with classical motion leads to
a detailed and intuitive understanding of these spectra and the corresponding
eigenstates.

In Sec.\ II we start with some general remarks on the relation between quantum
and classical description of the 2D electron motion in 1D lateral superlattices,
and we introduce suitable reduced units. In Sec.\ III we focus on the effect of
a simple harmonic magnetic modulation of arbitrary strength.  In Sec.\ IV we
include an electric modulation, which requires a somewhat different analytical
procedure. The inclusion of electric modulation seems also necessary from the
experimental point of view, since the ferromagnetic strips on the sample surface
introduce a periodic stress field in the sample, which acts as an electric
modulation on the 2DEG.  Finally, in Sec.\ V we summarize the essential features
derived in the paper and extend the discussion beyond the model of simple
harmonic modulations.

Some of the present results have been recently published in a preliminary
form.\cite{Manolescu98:PhysicaB}

\section{General remarks}
We consider a (non-interacting) 2DES in the $x$-$y$ plane subjected to a
magnetic field with $z$ component $B_z(x)=B_0+B_m(x)$ and an electrostatic field
in $x$ direction leading to a potential energy $U(x)$. Our aim is a close
comparison of the classical and the quantum description of the electron motion
(in terms of orbits and wavefunctions, respectively) in such fields, especially
in the case that $U(x)$ and $B_m(x)$ are periodic in $x$ with the same period
$a$ and vanishing average values.

To evidence the translation invariance in $y$ direction in the (either classical
or quantum) Hamiltonian
\begin{equation}
  H=\frac{1}{2m} \left( {\bf p} +e {\bf A} \right)^2 + U \, , \label{H_gen}
\end{equation}
we describe $B_z(x)$ by an $x$-dependent vector potential ${\bf
  A}(x)=A(x)\,\bf{e}_y$ with $A(x) =xB_0 + A_m(x)$ and $A_m(x)= \int_0^x dx'
B_m(x')$. Then $y$ is a cyclic variable and the canonical momentum $p_y$ is
conserved, and one obtains an (one-dimensional) effective Hamiltonian
$H(X_0)=p_x^2/2m + V(x;X_0)$. For $B_0 \neq 0$, the effective potential can be
written 
\begin{equation}
  V(x;X_0)= \frac{m}{2} \omega_0 ^2 \left(x-X_0 +\frac{A_m(x)}{B_0} \right)^2
  +U(x) \, , \label{Veff}
\end{equation}
where $X_0= -p_y /eB_0$ is the center coordinate of the effective potential and
$ \omega_0 = eB_0 /m$ is the cyclotron frequency, both in the absence of
modulation.

In the quantum description, the reduction to a one-dimensional problem is
achieved by the product ansatz $\Psi_{n,X_0}(x,y)=L_y^{-1/2} \exp (ip_y y/\hbar)
\psi_{n,X_0} (x)$ for the energy eigenfunctions, where $L_y$ is a normalization
length, and the discrete quantum number $n=0,1,2, \dots$ counts the nodes of the
reduced wavefunction $ \psi_{n,X_0} (x)$. If $U(x)$ and $A_m(x)$ are bounded,
the $\psi_{n,X_0} (x)$ drop Gaussian-like for $|x| \rightarrow \infty$, and for
a fixed value of the quasi-continuous quantum number $X_0$ the energy
spectrum $E_n(X_0)$ is discrete.

In the classical description, we use the equation $mv_y =p_y +e A(x)$, which may
also be derived directly from Newton's equation, to eliminate the velocity
$v_y$. The effective motion in $x$ direction is determined by $H(X_0)=E$.
Similar to the wavefunctions, the orbits for given constants of motion, $X_0$
and $E$, are bounded in $x$ direction, however the energy $E$ is a continuous
variable. For a given $E=E_{\mathrm{F}}$, each position $x$ (with
$U(x)<E_{\mathrm{F}}$) is the turning point of two orbits which are
characterized by the center coordinates
\cite{Menne98:1707} 
\begin{equation}
  X_0^{\pm}(x)=x+\frac{A_m(x)}{B_0} \pm R_0 \sqrt{1-\frac{U(x)}{E_{\mathrm{F}}}}
  \, ,
\label{X0pm} 
\end{equation}
obtained from $H(X_0)=E_{\mathrm{F}}$ for $v_x = p_x /m=0$. Here $R_0=
v_{\mathrm F} / \omega _0$ is the cyclotron radius of electrons moving with
energy $E_{\mathrm{F}}=mv_{\mathrm F}^2/2$ in the magnetic field $B_0$.  For
given $E_{\mathrm F}$ and $X_0$, orbits exist in intervals in which $X_0^-(x)
\leq X_0 \leq X_0^+(x)$ holds. This allows a convenient classification of the
possible orbits at fixed energy $E_{\mathrm F}$ and for varying
$X_0$.\cite{Menne98:1707} Of course, the same classification can also be done by
directly investigating the effective potential. This may be preferred if one is
interested in orbits at different energies but the same $X_0$.

The calculation of the orbits is a simple textbook problem, but must in general
be done numerically. In accordance with the translational symmetry of the
problem, we will in the following not distinguish orbits which differ only by a
rigid shift in $y$ direction.  If an electron is at time $t_i$ at position
$(x_i,y_i)$ on an orbit characterized by the constants of motion $E_{\mathrm F}$
and $X_0$, with turning points $x_l$ and $x_r$ ($x_l<x_i<x_r$), it moves towards
one of the turning points so that at time
\begin{equation}
  \label{timeintegral}
  t(x;X_0, E_{\mathrm F}) = t_i + \int_{x_i}^x \frac{dx'}{|v_x( x';
    X_0,E_{\mathrm F})|}
\end{equation}
it is at position $(x, y(x;X_0,E_{\mathrm F}))$, with
\begin{equation}
  \label{orbits}
  y(x;X_0,E_{\mathrm F})=y_i+ \int_{x_i}^x\frac{v_y( x'; X_0)}{v_x(
    x';X_0,E_{\mathrm F})}dx'\, ,
\end{equation}
where $|v_x(x; X_0,E_{\mathrm F})|=v_{\mathrm F} \sqrt{1-V(x;X_0)/E_{\mathrm F}}
=\omega_0 \sqrt{[X^+_0(x)-X_0][X_0-X^-_0(x)]}$ and $v_y(x; X_0)= (\omega_0/2)
[X^+_0(x)+ X^-_0(x) -2 X_0]$. If at one of the turning points $x_l$ or $x_r$,
where $v_x(x; X_0,E_{\mathrm F})=0$, the derivative $\partial V(x;X_0)/\partial
x$ vanishes, we call this turning point and this orbit ``critical''. At critical
turning points the integrals (\ref{timeintegral}) and (\ref{orbits}) diverge, so
that the critical orbits there asymptotically approach straight lines parallel
to the $y$ axis. For non-critical orbits the integrals (\ref{timeintegral}) and
(\ref{orbits}) converge as $x$ approaches the turning points, and the total
orbit can be composed out of right-running ($v_x >0$) and left-running ($v_x <
0$) pieces with finite traverse time $T(X_0,E_{\mathrm F})= \int_{x_l}^{x_r} dx
/ |v_x( x; X_0,E_{\mathrm F})| $.  The probability density of finding the
electron at position $x$ is $W(x;X_0,E_{\mathrm F})=1/[T(X_0,E_{\mathrm F})
|v_x( x; X_0,E_{\mathrm F})|] $. This is the classical analog to 
$|\psi _{n,X_0} (x)|^2$.

If $U(x)=U(x+a)$ and $B_m(x)=B_m(x+a)$ are periodic with period $a$, as we will
assume in the following, the effective potential, Eq.~(\ref{Veff}), has the
symmetry $V(x+a;X_0+a)=V(x;X_0)$. As a consequence, the energy spectrum is also
periodic, $E_n(X_0+a)=E_n(X_0)$, and can be restricted to the ``first Brillouin
zone'' $0 \leq X_0 \leq a$. The eigenfunctions can be taken to satisfy
$\psi_{n,X_0+a} (x)=\psi_{n,X_0} (x-a)$. The corresponding classical symmetry is
that an orbit characterized by $E_{\mathrm F}$ and $X_0+a$ differs from that
characterized by $E_{\mathrm F}$ and $X_0$ only by a rigid shift of amount $a$
in $x$ direction.

The dispersion of the energy bands $E_n(X_0)$ implies a group velocity in $y$
direction, 
\begin{equation}
  \label{velocity}
  \langle n, X_0 | v_y |n, X_0 \rangle = - \frac{1}{m \omega_0} \frac{d
    E_n(X_0)}{ dX_0} \, ,
\end{equation}
which is the expectation value of the velocity operator in the energy eigenstate
$\psi_{n,X_0} $. It is the quantum equivalent to the classical drift velocity,
i.e.\ the average velocity (in $y$ direction) along the corresponding classical
orbit. The drift velocity in $x$ direction vanishes, since the orbits are
bounded in $x$ direction.

\subsection{Suitable units}
For an economic comparison of classical and quantum aspects it is important to
use suitable length and energy units, which are meaningful for both the
quantum description and the classical limit. Doing so, we will see that  the
classical features 
depend on fewer scaled parameters than the quantum ones. To be specific but
still rather general, we assume in the following periodic modulations of the
form $B_m(x)=B_m^0 \, b(Kx)$ and $U(x)=V_0 \, u(Kx)$ for the magnetic and the
electric modulation, respectively, where $b(\xi)$ and $ u(\xi)$ are
dimensionless periodic functions with period $2\pi$ and vanishing average
values. Thus $B_m(x)$ and $U(x)$ have the same period $a=2\pi /K$, but may have
different shapes and phases.  In the numerical examples we will use for both
$b(\xi)$ and $u(\xi)$ simple cosines, eventually with a phase shift.

The average magnetic field $B_0$ sets, with the magnetic length $l_0 =
\sqrt{\hbar /(m \omega_0)}$ and the cyclotron energy $\hbar\omega_0$, both a
length 
and an energy scale, which are useful for quantum calculations, but have no
meaning for the classical motion. For the discussion of commensurability
effects, such as the Weiss oscillations, the cyclotron orbits must be
compared with the period $a$ of the modulation. Therefore $a$ is a natural
choice for the lengths unit. The choice of a suitable energy unit is motivated
as follows.

Classically, $B_0$ determines only the
cyclotron frequency $\omega_0$, and one needs an independent length $l$ to
define an energy scale $ V_{\rm  mag}= m \omega_0 ^2 l^2 /2$. Using $l$ as
length unit, we may define dimensionless variables $\xi = x/l$ and $\xi_0 =X_0
/l$. 
%
 The effective potential, Eq.~(\ref{Veff}), then can be
written as $ V(x;X_0)= V_{\rm mag} \, \tilde{v}(\xi ; \xi_0)$ with 
\begin{equation}
  \tilde{v}(\xi;\xi_0) = [\xi-\xi_0 + s \, a(Kl \xi)/Kl ]^2 + w \, u(Kl \xi) \,
  , \label{Veffxi}
\end{equation}
where $s = B_m^0 /B_0$, $a(\zeta)=\int_0^{\zeta} d \zeta ' b( \zeta ')$, and
$w=V_0 / V_{\rm mag}$. In the quantum description, the kinetic energy operator
$-(\hbar ^2/2m) d^2/dx^2 = - E_l\, d^2/d\xi ^2$, introduces a new energy
scale $E_l = \hbar ^2 /( 2m l^2) $, which has no classical analog.
Introducing the energy ratio $ \alpha = E_l /V_{\rm mag}$, we write
the effective Schr{\"o}dinger equation  as
\begin{equation}
\left[- \alpha \,
\frac{d^2}{d \xi ^2} + \tilde{v}(\xi ; \xi_0) - \tilde{\varepsilon}_n
  (\xi_0) \right] \, \tilde{\psi}_{n, \xi _0} (\xi) =0 \, ,
  \label{schroedinger} 
\end{equation}
with  $\tilde{\varepsilon}_n(\xi_0)\! = \!
E_n(X_0)/V_{\rm  mag}$ and $ \tilde{\psi}_{n, \xi _0} (\xi) \! = \!
\sqrt{l}\, \psi _{n, X_0} (x)$.

If we would take $l=l_0$, we had $ V_{\rm mag}= E_l= \hbar \omega_0/2$ and thus
simply $ \alpha =1$. The effective potential Eq.~(\ref{Veffxi}) would then
depend on the constant of 
motion $\xi_0$ and, in addition, on three dimensionless
model parameters, $s$, $w$, and $Kl_0$.  To specify an eigenstate or, in the
classical description, a trajectory, one further needs an energy value
$\tilde{\varepsilon}$ as a second constant of motion. 
   A description that, for fixed
  constants of motion, needs {\em three} parameters to specify the effective
  potential and, furthermore, relies on $l_0$ and $ \hbar \omega_0$, which have
  no meaning in classical mechanics,  is rather clumsy and  not acceptable.

Instead we take $l=1/K$ and, therefore, $ V_{\rm mag}= V_{\rm cyc}$,
where  $ V_{\rm cyc} =m \omega_0^2 /(2 K^2)$ is the energy of a classical
cyclotron orbit of radius $1/K$ in the homogeneous magnetic field $B_0$.
  Now the effective potential Eq.~(\ref{Veffxi}) depends only on the {\em two}
  dimensionless modulation strengths $s$ and $w =V_0 / V_{\rm cyc} $, which
  both are well defined within the classical approach. Also the
 constants of motion, $\xi_0= KX_0$ and $\tilde{\varepsilon}= E/ V_{\rm
  cyc} =(K R_0)^2$, including the dimensionless version of
Eq.~(\ref{X0pm}), 
\begin{equation}
  \xi _0 ^{\pm} (\xi) = \xi + s \, a( \xi) \pm\sqrt{\tilde{\varepsilon}- w \,
    u(\xi)} \, , \label{xi0pm}
\end{equation}
remain meaningful in the classical limit.
This choice of units will also be very useful for a systematic discussion of
the quantum mechanical energy spectra. Quantum mechanics enters the effective
Schr{\"o}dinger equation (\ref{schroedinger}) only via the parameter $\alpha =
(l_0 K)^4$, which scales the kinetic energy. It determines the only true
quantum 
aspect of the spectrum, namely the spacing of the energy levels
$\tilde{\varepsilon}_n (\xi_0)$. We will see in Sec.~\ref{starkemod} that all
the essential structural features of the energy spectrum, e.g. the complicated
back-folded structure due to the coexistence of ``channeled'' and ``drifting''
states, are determined  solely by the ``classical'' parameters $s$ and $w$. The
density of the quantized levels $\tilde{\varepsilon}_n (\xi_0)$, on the other
hand, increases with increasing ratio $a/l_0$. 

As a simple example one may
consider the well known case of a weak electric or magnetic cosine modulation,
which leads to modified Landau bands of oscillatory
width.\cite{Gerhardts89:1173,Zhang90:12850,Vasilopoulos90:393,Peeters93:1466}
The band width assumes minima near the ``flat band'' energies
$E_{\lambda}^{\pm} =m(\omega_0 a)^2 (\lambda \pm 1/4)/8$, with  ``$+$''
(``$-$'')  for
magnetic (electric)  modulation and  $\lambda =1,~ 2,~ \dots$.
These flat band energies are distinct multiples of our energy unit $V_{\rm
  cyc}$, and occur at  $\tilde{\varepsilon}_{\lambda}^{\pm} =
\pi ^2 (\lambda \pm 1/4)$, independent of the special values of the model
parameters $B_0$ and $a$. The level spacing, on the other hand, is of the order
$\hbar \omega_0$ and depends in our units on $\hbar \omega_0 / V_{\rm cyc} =2
l_0^2 K^2= 2\sqrt{\alpha}$.

\section{Magnetic cosine modulation} 
We first consider a pure magnetic modulation, $U(x) \equiv 0$, $B_m(x)=B_m^0
\,b( Kx)$, so that the effective potential Eq.~(\ref{Veffxi})
becomes
\begin{equation}
  V(\xi;\xi_0) = V_{\rm cyc} [\xi-\xi_0 + s \, a( \xi) ]^2 \,.
  \label{Veffxi_mag}
\end{equation}
For $s=0$ one obtains the well known Landau levels and the Landau oscillator
wave functions, $f_{nX_0}(x)$. We use the set $f_{nX_0}$ as the basis of our
Hilbert space in order to obtain numerical solutions for $s\neq 0$, by numerical
diagonalization of $H(X_0)$.  The electron effective mass is that of GaAs,
$m=0.067m_0$. We further assume spin degeneracy. For the numerical parameters
chosen here the size of the basis will vary between 150-300 Landau levels.

Before discussing the numerical results we summarize some properties of the 
effective potential and of Eq.~(\ref{xi0pm}), which now reduces to
\begin{equation}
  \xi_0^{\pm}(\xi) = \xi + s \, a( \xi) \pm K R_0 \,. \label{xi0pm_m}
\end{equation}
For a fixed $\xi_0$ the local extrema of the effective potential, given by
$\partial V(\xi,\xi_0)/ \partial \xi=0$, are the points
where the total magnetic field is zero, i.~e. the roots of 
\begin{equation}
  1+s \, b(\xi)=0\,,
\label{der1} 
\end{equation} 
and the points where the effective potential is zero, i.~e. the roots of
\begin{equation}
\xi-\xi_0+s\, a(\xi)=0\,. 
\label{der2} 
\end{equation} 
An important aspect for the following discussion is that the roots of the first
kind, Eq.(\ref{der1}), if existent, are independent of $\xi_0$, while those of
the second kind, Eq.(\ref{der2}), do depend on $\xi_0$.  We will see that orbits
with $\xi$ values near roots of the first kind are channeled, while those with
$\xi$ values near roots of the second kind
are drifting orbits.
The analytic dependence of the effective potential on the relevant position
coordinate $\xi$ is determined by the modulation strength $s$.  Therefore the
number of its possible zeroes, the classification of orbits and the energy
spectrum depend critically on the parameter $s$. To demonstrate this, we choose
in the following examples $b( \xi) = \cos \xi$, and consequently 
$a( \xi)= \sin \xi$.

\subsection{Weak modulation, $s \leq 1$}
For $s < 1$, the effective potential has exactly one minimum of the second kind
for each value $\xi_0$, which is due to the confinement by the average magnetic
field.  The functions $\xi_0^{\pm}(\xi)$ in equation (\ref{xi0pm_m}) have no
extrema.  For each value $\xi_0$ they determine exactly one orbit, which is a
drifting cyclotron orbit. By this we mean a self-intersecting orbit consisting
of loops along each of which the azimuth angle in velocity space, $ \varphi =
\arctan (v_y /v_x)$ increases by $ 2\pi $.  A typical example is illustrated in
Fig.~\ref{illu1} for $s=0.5$, $\xi_0=\pi/2$ (i.e.\ $X_0=a/4$), and two energy
values $E_{\mathrm F}$.  Figure~\ref{illu1}(a) shows the effective potential.
For a given energy $E=E_{\mathrm F}$ (horizontal line) a classical orbit exists
where $V(\xi;\xi_0) \leq E_{\mathrm F}$.  Figure~\ref{illu1}(b) shows the
location of the turning points as the crossing points of the horizontal line
$\xi_0=\pi/2$ with the functions $\xi_0^{\pm}(\xi)$. The corresponding drifting
orbit exists in the interval with $\xi_0^-(\xi) \leq \xi_0 \leq \xi_0^+(\xi)$.
The orbits in
real space are illustrated in  Fig.~\ref{illu1}(c). 
In Fig.~\ref{illu2} we plot the corresponding quantities for $s=0.5$ and
$\xi_0=\pi$ (i.e.\ $X_0=a/2$). In this case the effective potential is symmetric
with respect to the center coordinate $X_0$. As a consequence, the orbits are
closed and their drift velocity in $y$ direction is zero.

For small energies, $ E_{\mathrm F} /V_{\mathrm cyc} =(KR_0)^2 < \pi ^2$ (i.e.
$2 R_0 < a$), the extents of the orbits in $x$ direction are smaller than a
modulation period and essentially determined by the local values of the total
magnetic field. At high energies, $ E_{\mathrm F} /V_{\mathrm cyc} \gg 1$, the
orbits extend over several periods of the modulation and the extent of an orbit,
i.e., the width of the effective potential valley at the corresponding energy,
is determined by the cyclotron radius in the average magnetic field ($x_r - x_l
\approx 2 R_0$).

In Fig.~\ref{figure1}(a) we display the first 50 energy bands $E_{n\xi_0}$
calculated from the (first 150) original, degenerated Landau levels, for s=0.5.
The level spacing of the lowest energy bands is seen to follow the local value
of the total magnetic field, Fig.~\ref{figure1}(b). This is expected from the
local approximation $E_{n\xi_0}\approx(n+1/2)\hbar e B(\xi_0)/m$, which is valid
if the extent of the wavefunctions $\psi_{n,X_0}(x)$ is smaller than the
modulation period. With our energy unit $V_{\mathrm cyc}$ the apparent level
spacing of energies which are independent of the period $a$ becomes proportional
to $\sqrt{\alpha}$. For example, if the local approximation
$E_{n\xi_0}\approx(n+1/2)\hbar \omega(\xi_0)$ holds for $E_{n\xi_0} < 4
V_{\mathrm cyc}$, as in Fig.~\ref{figure1}(a), this implies that it holds for
$n+1/2 < 4V_{\mathrm cyc}/[\hbar \omega(\xi_0)]=2[\omega_0 / \omega(\xi_0)] /
\sqrt{\alpha}$.  Thus, the number of bands which are well described by the local
approximation increases quadratically with increasing modulation period $a$.

The local approximation fails at higher energies when the width of the
wavefunctions becomes larger than the period of the modulation, and the
structure of the energy spectrum changes. Indeed it is well known from the limit
of very weak magnetic modulation, $s \ll 1$, that in contrast to this local
approximation the bands become flat at the energies $E_{\lambda} / V_{\mathrm
  cyc}= \pi ^2 (\lambda+1/4)$, for $\lambda
=1,~2,\dots$.\cite{Vasilopoulos90:393,Peeters93:1466,Ye95:3013} These flat band
conditions are the quantum equivalents to the classical commensurability
conditions leading to the Weiss oscillations in magnetotransport, and do not
change their positions in a plot like Fig.~\ref{figure1}(a), even if we change
the modulation period.  A larger modulation period $a$ just leads to a higher
density of the energy bands.

In Fig.~\ref{figure1}(c) we plot for $\xi_0=\pi/2$ the effective potential and
the square of the energy eigenfunctions for the energy values considered in
Fig.~\ref{illu1}. Width and location of the wavefunctions in the effective
potential is in close agreement with that of the corresponding classical orbits.
In Fig.~\ref{figure1}(d) we plot the corresponding quantities for the symmetric
situation $\xi_0=\pi$, to be compared with Fig.~\ref{illu2}. These wavefunctions
belong to (relative) extrema of the energy bands, and thus have zero group
velocity, in agreement with the zero drift velocity of the corresponding
classical orbits. The wavefunctions in Fig.~\ref{figure1}(c) belong to finite
energy dispersion and describe motion in the positive ($n$=3) and the negative
($n$=43) $y$ direction, respectively, in agreement with the correponding
classical orbits in Fig.~\ref{illu1}. For large quantum numbers $n$ and weak
modulation the group velocities can be shown to reduce quantitatively to the
drift velocities of the corresponding classical orbits.

In Fig.~\ref{figure2} we consider the ``critical'' situation $s=1$.  The
derivatives $\xi_0^{\pm\prime}(\xi_{ex})=0$ and
$\xi_0^{\pm\prime\prime}(\xi_{ex})=0$ vanish for $\xi_{ex} = (2 p +1)\pi$ ($p$
integer), i.e.\ for the positions where the magnetic field vanishes,
Eq.(\ref{der1}).  For all values of $\xi_0$ the effective potential
(\ref{Veffxi}) becomes flat at these points $\xi_{ex}$ (see
Fig.~\ref{figure2}(c)).  The classical situation is as for $s<1$ with the
exception that for $\xi_0 = \xi_{ex} \pm KR_0$ there are ``critical'' orbits
which asymptotically approach straight lines parallel to the $y$ axis on 
their left (for $+$) or their right (for $-$) side, where $B(x)=0$. 
The dashed lines plotted over the energy spectrum of Fig.~\ref{figure2}(a) show
the evolution of the flat regions of the effective potential with $\xi_0$, i.e.\ 
the parabolas resulting from $V(\xi_{ex};\xi_0)$ with $p=0$ and $p=\pm 1$. In
the first Brillouin zone these lines are seen as the back-folding of the lowest
parabola centered on $\xi_{ex}$ with $p=0$, and they are an indication of a kind
of a free electron motion along the lines where the magnetic field is zero.
Close to these parabolas the energy bands have large dispersion near inflexion
points, and the energy separation between adjacent bands is minimum. Similar
features have been obtained in the energy spectra for single magnetic wells by
Peeters and Matulis.\cite{Peeters93:15166} In other words, such states
experience a weak effective magnetic field, due to the constant effective
potential over a substantial spatial region. The wavefunctions corresponding to
states with large energy dispersion have large amplitudes at the positions of
flat effective potential (vanishing total magnetic field).  This is demonstrated
for two selected states [($n$=43, $\xi_0$=3.18) and ($n$=5, $\xi_0$=0.50)] in
Fig.~\ref{figure2}(b), together with the probability distributions of the
corresponding classical orbits.  The effective potentials together with the
corresponding classical orbits are plotted in Fig.~\ref{figure2}(c). The
trajectory corresponding to the state ($n$=5, $\xi_0$=0.50) is close to a
critical orbit with a critical right turning point.  This leads to an enhanced
probability density near that point, which is also reflected in the quantum
mechanical probability density.

We will see that for slightly stronger modulation a new type of nearly free
motion occurs with energies close to the parabolas $V(\xi_{ex};\xi_0)$ in the
energy spectrum.

\subsection{Strong modulation, $ s > 1$ }
\label{starkemod}
For $s>1$, $\xi_0^{\pm\prime}(\xi)=0$ at $a(\xi) \equiv \cos \xi =-1/s$, and
$\xi_0^{\pm}(\xi)$  has extrema  at the following positions:
\begin{eqnarray}
{\mathrm{minima:}}~~\,  \xi_{min}^{(p)} &=& (2p+1)\pi + \delta,  \nonumber \\
{\mathrm{maxima:}}~~  \xi_{max}^{(p)} &=& (2p+1)\pi - \delta,  \label{xiextren}
\end{eqnarray}
where $p$ is an integer and $\delta = \arctan \sqrt{s^2-1}$. The values at these
extrema are
\begin{eqnarray}
  \xi_0^{\pm}(\xi_{min}^{(p)})&=&(2p+1)\pi - g(s)\pm KR_0 \nonumber \\ 
  \xi_0^{\pm}(\xi_{max}^{(p)})&=&(2p+1)\pi + g(s)\pm KR_0 \, , \label{fextrem}
\end{eqnarray}
where 
\begin{equation}
\label{gvons}
g(s)= \sqrt{s^2 -1} -\arctan \sqrt{s^2-1} >0 \, .
\end{equation}
The effective potential $V(\xi;\xi_0)$ has extrema of the first kind,
Eq.~(\ref{der1}), at the same positions. The extrema with values $V(
\xi_{min}^{(p)};\xi_0) =V_{\rm cyc} [ (2p+1)\pi - g(s) -\xi_0]^2$ are minima if
$ (2p+1)\pi >\xi_0$, and maxima otherwise, and those with values $V(
\xi_{max}^{(p)};\xi_0) =V_{\rm cyc} [ (2p+1)\pi + g(s) -\xi_0]^2$ are maxima
if $ (2p+1)\pi >\xi_0$, and minima otherwise.
\subsubsection{Classical approach}
The number of zeroes of $\xi_0^{\pm}(\xi)-\xi_0$ depends on both $s$ and
$\xi_0$.  If $g(s)< \pi$, $\xi_0^{\pm}(\xi)-\xi_0$ has at most three zeroes.  If
$\xi_0=\xi_0^{\pm}(\hat{\xi})$ for any $\hat{\xi}$ satisfying $\xi_{max}^{(p)} <
\hat{\xi}< \xi_{min}^{(p)}$, i.e. if $(2p+1)\pi -g(s) < \xi_0 \mp KR_0
<(2p+1)\pi+g(s)$, $\xi_0^{\pm}(\xi)-\xi_0$ has three zeroes. The same argument
holds for Eq.~(\ref{der2}), i.e.\ the effective potential has three zeroes.  For
$(2p-1)\pi +g(s) < \xi_0 \mp KR_0 < (2p+1)\pi -g(s)$, on the other hand, there
exists only a single zero.

In Fig.~\ref{illu3} we show, for $s=2$ [ i.e.\ $g(s)$=0.685], an example where
the effective potential has a single zero near $\xi/2\pi=0.1$, so that for
sufficiently low energy only a single drifting orbit exists.  The number and the
type of the possible orbits depend on the energy.  At the highest energy shown
in Fig.~\ref{illu3}(a) two orbits exist (solid lines in Fig.~\ref{illu3}(d)).
There is a drifting cyclotron orbit extending over more than two periods of the
modulation, with the left turning point on $\xi_0^+(\xi)$ (uppermost curve in
Fig.~\ref{illu3}(c)) near $\xi/2\pi=-1.1$, and the right turning point on
$\xi_0^-(\xi)$ (bottom curve in Fig.~\ref{illu3}(c)) near $\xi/2\pi=1.3$. Near
the relative minimum of $\xi_0^-(\xi)$ close to $\xi/2\pi=1.7$, which
corresponds to a relative minimum of the effective potential (thick dashed line
in Fig.~\ref{illu3}(a)), there exists a ``channeled orbit'' moving in positive
$y$ direction.  We define channeled orbits as trajectories which have both
turning points either on $\xi_0^+(\xi)$ or on $\xi_0^-(\xi)$, in contrast to the
drifting orbits with one turning point on $\xi_0^+(\xi)$ and the other on
$\xi_0^-(\xi)$. In contrast to the self-intersecting drifting orbits, the
channeled orbits are always confined to less than a single modulation period,
and they move without self-intersections in a relatively narrow interval of
angles around the positive or the negative $y$ direction [see
Fig.~\ref{illu3}(d)]. Note that the curvature of the trajectories changes sign
at the positions where the total magnetic field vanishes, see
Fig.~\ref{illu3}(b).

If we lower the energy to $E/V_{\rm cyc}=40$, we arrive in Fig.~\ref{illu3} at a
situation where only a single drifting orbit exists (dashed lines). In general,
the extent in $\xi$ direction of the drifting orbits decreases with decreasing
energy. At the lowest energy indicated in Fig.~\ref{illu3} (lowest dotted line
in (a) and innermost lines in (c)), we have again a drifting orbit near
$\xi/2\pi=0.1$ and a channeled orbit around $\xi/2\pi=0.6$. At this low energy,
the extent of the drifting orbit is considerably smaller than the modulation
period.

In Fig.~\ref{illu4} we show, for the same modulation strength, $s=2$, a
situation, $\xi_0=\pi$, where the effective potential has three zeroes, as is
emphasized in the inset of Fig.~\ref{illu4}(a). These zeroes are separated by
two shallow maxima. If a (positive) $E$ value below these maxima is chosen, one
finds three narrow drifting cyclotron orbits located around the zeroes of the
effective potential (solid lines). For higher energies one may find either one
drifting and two channeled orbits (dotted lines) or a single drifting orbit
(e.g.\ for $0.5<E/V_{\mathrm cyc}<30$, not indicated in the figure). Actually
the ``drifting'' orbits located around $\xi=\pi$ have zero drift velocity due to
symmetry reasons.

In summary, for $0<g(s)<\pi$ we find for given values of the constants of
motion, $\xi_0$ and $E$, at least one and at most three orbits.  For larger
values of $s=B_m^0/B_0$, more orbits may exist for a given pair of $\xi_0$ and
$E$ values. A careful analysis of the extrema of the functions
$\xi_0^{\pm}(\xi)$ shows, e.g., that for $\pi < g(s) < 2\pi$ between three and
five orbits belong to the same pair of $\xi_0$ and $E$.  We will come back to
this case below.

Apparently the plots of the effective potential $V(\xi;\xi_0)$ are very useful
to see which orbits are possible for a fixed value of the center coordinate
$\xi_0$ and different energies. Channeled orbits exist in side valleys near
relative minima of $V(\xi;\xi_0)$.  If, on the other hand, the energy of the
motion is given, the plots of the locations of turning points $\xi_0^{\pm}(\xi)$
is very useful to classify the possible orbits for different values of $\xi_0$.
Channeled orbits exist near relative minima of $\xi_0^-(\xi)$ and relative
maxima of $\xi_0^+(\xi)$.

\subsubsection{Quantum calculation}
The energy spectra become more complicated in the case $s>1$,
Fig.~\ref{s_2_spec} and \ref{s_5_spec}. Regions of different character can be
distinguished in these spectra.  Areas, where the energy bands are nearly
parallel lines with low dispersion (region I), alternate with regions, where
steep bands with large dispersion seem to cross bands with weak dispersion
(region II). In fact the energy bands never cross each other and the apparent
intersections are anti-crossing points with exponetially small gaps.

The boundaries of these regions are given by classical values only. If the
energy is scaled by the classical cyclotron energy $V_{\mathrm cyc}$, for fixed
modulation strength $s$ the regions II are surrounded by the parabolas
$E=V(\xi_{\mathrm min}^{(p)};\xi_0 )$ and $E=V(\xi_{\mathrm max}^{(p)};\xi_0 )$
which, for $\mid p \mid \le 2$, are indicated by dashed lines in the spectra.
For any fixed $\xi_0$ such a pair of parabolas gives the minimum and the maximum
value of a certain side valley of the effective potential $V(\xi;\xi_0)$
(extrema of the first kind, see Eq.~(\ref{der1})).  The energy interval in
between these values indicates the depth of that valley, i.e.\ an energy range
in which classically channeled orbits exist, in addition to the drifting orbits.

In Fig.~\ref{s_2_spec}(b) the effective potential is plotted for the symmetric
case $\xi_0=\pi$ (dotted line), corresponding to the classical situation
described in Fig.~\ref{illu4}.  Also shown are the states for $n=0$ (lower solid
line) and for $n=20$ (upper solid line) and $n=22$ (upper dashed line).
Apparently, state $n=20$ corresponds to a classical drifting orbit, whereas
$n=22$ is the symmetric superposition of two states corresponding to channeled
orbits in the side valleys. The latter has practically the same energy as the
corresponding antisymmetric superposition ($n=21$), which is not shown. On the
scale of Fig.~\ref{s_2_spec}(a), all states $n=20$, 21, and 22 seem to have the
same energy, $E/V_{\rm cyc} \approx 38$. The states $n=21$ and 22 are
hybridizations of states belonging to the branches with high energy
dispersion and opposite sign of the group velocity
\begin{equation}
\langle v_y \rangle = - \frac{K}{m\omega_0}\frac{d E_{n\xi_0}}{d\xi_0} \,.
\label{vy}
\end{equation}
Figure~\ref{s_2_spec}(c) shows the effective potential for the asymmetric case
$\xi_0/2\pi =0.2$, corresponding to the classical situation described in
Fig.~\ref{illu3}. Here we show six states, the ground state $n=0$ located near
the zero of the effective potential, the two ``channeled'' states ``bound'' in
the potential valley around $\xi/2\pi =0.6$, the extended ``drifting'' state
$n=25$ near $E=40 V_{\rm cyc}$, the extended state $n=44$, and the localized
``channeled'' state $n=45$. The energies of all these states are indicated in
Fig.\ref{s_2_spec}(a). The states which extend over more than a period of the
modulation belong to weakly dispersive energy bands and correspond to the
classical drifting orbits. The states belonging to the energy branches with
strong dispersion have large amplitudes in side valleys of the effective
potential and vanish practically outside these valleys. They correspond to
classical channeled orbits. The apparent number of nodes of the large-amplitude
parts of these ``channeled'' states increases with energy as if they were truly
bound states in these narrow valleys. Note, however, that the wave functions of
channeled states still have $n$ nodes, but the corresponding oscillations are
not observable at the scale of the figure. Outside the valleys, the maxima in
between the nodes are a few orders of magnitude smaller than the main peaks
inside the valleys.

The number of quantized states within a given valley of the effective potential
depends on the average magnetic field and the period of the magnetic modulation,
even if the parameters $V_{\rm cyc}$ and $s$ are fixed. In Fig.\ref{s_2_spec}
the period $a$, or the field $B_0$, is too small to have states quantized in the
low-energy triple minimum of the effective potential for the symmetric case of
Fig.\ref{s_2_spec}(b) (see also Fig.\ref{illu4}). To investigate this situation,
we show in Fig.\ref{s_5_spec}(a) a denser spectrum for the same modulations
strength $s=2$. In Fig.\ref{s_5_spec}(b) the spectrum near $\xi_0=\pi$ is
enlarged. Five energy values are indicated, and in Fig.\ref{s_5_spec}(c) the
corresponding (squares of the) wavefunctions are plotted for the states $n=0$,
2, 3, 4, and 5, together with the effective potential.  The antisymmetric state
$n=1$, which is nearly degenerate with $n=2$, is not shown.  This demonstrates
that all the classical features have their quantum analog, provided the model
parameters (here $\alpha$) are suitably chosen.

For the magnetic cosine modulation, the depth of the valleys of the effective
potential,
\begin{equation}
  | V(\xi_{\mathrm max}^{(p)};\xi_0 ) - V(\xi_{\mathrm min}^{(p)};\xi_0 ) | /
  V_{\mathrm cyc} = 4 g(s) | (2 p +1)\pi - \xi_0| \, , \label{stripwidth}
\end{equation}
increases with the energy (i.e.\ with $|p|$ for fixed $ \xi_0$), and thus more
and more channeled states appear at higher energies.  For sufficiently high
energies the strips with channeled states in the energy spectra may thus extend
over the whole Brillouin zone. This will also happen for sufficiently large $s$.
The energy dispersion of the channeled states depends strongly, nearly
quadratically, on $\xi_0$ according to Eq.(\ref{Veffxi_mag}), which expresses
the nearly free motion of the electrons on channeled orbits in $y$ direction.

For $s>1$ and $g(s)<\pi$, the area of the regions II of the spectrum increases
with increasing $s$, and the area of the regions I shrinks accordingly. For
$g(s)= \pi$, one has $ V(\xi_{\mathrm min}^{(p)};\xi_0 ) = V(\xi_{\mathrm
  max}^{(p-1)};\xi_0 )$ and the corresponding parabolas coincide, leaving no
room for regions I. If the modulation is so large that $g(s) \geq \pi$, drifting
and channeled states coexist everywhere in the spectrum.  In Fig.~\ref{s_6_spec}
we have chosen $s=5$, corresponding to $g(s)=3.53$.  Close to the edges of the
Brillouin zone, e.g. for $\xi_0/2\pi=0.016$, Fig.~\ref{s_6_spec}(c), we can
identify channeled, e.g.\ $n=22$ and $n=16$, and drifting states, e.g.\ $n=19$.
But now these drifting states are relatively narrow, confined in local minima of
the effective potential and not in the wide potential well centered around
$\xi_0$, which is given by the confinement due to the average magnetic field.
This case is already known from the discussion of Fig.~\ref{illu4}.  The local
minimum of the effective potential at $\xi=0$ is a minimum of the second kind,
with vanishing potential.  Consequently, these drifting states are similar to
weakly perturbed Landau levels with energy gaps $\hbar eB(\xi_0)/m$, as can be
observed by a careful look at Fig.~\ref{s_6_spec}(a).  In the center of the
Brillouin zone say for $\xi_0/2\pi=0.493$, Fig.~\ref{s_6_spec}(d), the effective
potential has three zeroes of the second kind (see Eq.(\ref{der2})) near
$\xi=\xi_0$.  We therefore can find similar narrow drifting states, like $n=5$
and $n=6$, but also wide drifting states at higher energies, like $n=17$ and
channeled states in local minima of first kind, like $n=7$.  As in the classical
picture, the velocity of these narrow drifting states is in general lower than
that of the channeled states.

\section{Mixed harmonic modulations}
For sufficiently strong mixed electric and magnetic modulations one expects a
similar situation as for the strong magnetic modulation, with a coexistence of
channeled and drifting orbits, and their quantum analogs.  In the presence of an
electric modulation, we have no explicit analytic expressions for the minima of
the effective potential, Eq.~(\ref{Veffxi}), not even for simply harmonic
modulations. Nevertheless, a qualitative understanding of the classical and the
corresponding quantum mechanical motion is possible.  For a given constant of
motion $\xi_0$ the effective potential has side valleys with possible channeled
orbits, if $\partial V(\xi ;\xi_0)/ \partial
\xi =0$ has more than one solution $\xi$. This is the case, if the function
\begin{equation}
  \xi_0 (\xi) = \xi + s \, a(\xi) -\frac{w}{2}\, \frac{u' (\xi)}{1+s \, b( \xi)}
  \, , \label{pardarst1}
\end{equation}
with $w=V_0/ V_{\rm cyc}$,
assumes the value $\xi_0 (\xi)= \xi_0 $ at more than one $\xi$ value. At such
$\xi$ values the effective potential has extrema with the values
\begin{equation}
  V(\xi ;\xi_0(\xi))/ V_{\rm cyc} = w \, u (\xi) + \left[ \frac{w}{2}\, \frac{u'
    (\xi)}{1+s \, b( \xi)} \right] ^2 \, .
  \label{pardarst2} 
\end{equation}
To be specific, we choose for the following $b(\xi)=\cos \xi$, $a(\xi)=\sin \xi$
and $u (\xi)=\cos( \xi+\varphi_s) $.

Apparently, Eqs.~(\ref{pardarst1}) and (\ref{pardarst2}) provide a parametric
representation of the possible relative extrema of the effective potential in
the energy-versus-$\xi_0$ diagram, similar to the dashed lines in
Figs.~\ref{s_2_spec} and \ref{s_5_spec} which define the regions II where
``channeled'' states coexist with ``drifting'' ones. Due to the symmetries
$\xi_0(\xi +2\pi)=\xi_0(\xi)+2\pi$ and $V(\xi+2\pi ;\xi_0(\xi+2\pi)) =V(\xi
;\xi_0(\xi))$, it is sufficient to consider only one period $0 \leq \xi \leq
2\pi$ of the parameter $\xi$, provided the $\xi_0(\xi)$ values are back-folded
into the first Brillouin zone.

For strong magnetic modulation ($s>1$) the denominators in
Eqs.~(\ref{pardarst1}) and (\ref{pardarst2}) lead to poles. Then the regions of
type II extend to arbitrary high energies, similar to the case of pure magnetic
modulation. A qualitatively different behavior is obtained for weak magnetic,
but arbitrarily strong electric modulation, since then the denominators of
Eqs.~(\ref{pardarst1}) and (\ref{pardarst2}) remain positive (for $s<1$).
Consequently, for given values of $w$, $s$, and $\varphi_s$ the possible values
of $V(\xi ;\xi_0(\xi))$, Eq.~(\ref{pardarst2}), are bound
and channeled orbits can exist only below a certain energy. 
\subsection{Pure electric modulation}
As a particularly simple example we consider a pure electric cosine modulation,
$s=0$, $\varphi_s=0$.  For weak modulation ($w<2$) the function $\xi_0(\xi)$ of
Eq.~(\ref{pardarst1}) has a unique inverse, i.e.\ the effective potential $V(\xi
;\xi_0)$ has for all values of $\xi_0$ only a single extremum, namely the
absolute minimum, and no ``channeled'' states should be expected. The energy
spectra for this weak-modulation limit are well known
\cite{Gerhardts89:1173,Winkler89:1177,Shi96:12990} and will not be reproduced
here. Apart from a phase shift, they look similar to Fig.~\ref{figure1}(a) but
with flat bands near $E/V_{\mathrm cyc} =\pi ^2 (\lambda -1/4)$, for $ \lambda =
1, ~2, \dots$.

The corresponding classical trajectories at sufficiently high energies are
drifting cyclotron orbits. At very low energies, $E<V_0=w V_{\mathrm cyc}$, a
peculiarity occurs, since then the classical trajectories are captured within a
single valley of the electric potential, with turning points given by
Eq.~(\ref{xi0pm}) in the interval $\xi_c \leq \xi \leq 2\pi - \xi_c$ (modulo
$2\pi$) with $\xi_c =\arccos (E/V_0) >0$. For $\xi_0$ values in the interval
$\xi_c < \xi_0 < 2\pi - \xi_c$ these trajectories are self-intersecting drifting
orbits, whereas for $\xi_0<\xi_c $ and $\xi_0 > 2\pi - \xi_c$ there exist
channeled orbits with $v_y > 0$ and $v_y <0$, respectively. The orbit with
$\xi_0 = \xi_c $ approaches the left turning point at $\xi = \xi_c $ with a
tangent parallel to the $x$ axis, and that with $\xi_0=2\pi -\xi_c$ does the
same at the right turning point $\xi=2\pi - \xi_c$.  This peculiar low-energy
behavior is, of course, not restricted to the weak modulation limit, but occurs
always when the trajectories are captured in a minimum of the electric
potential, i.e.\ for $E<V_0$. It is demonstrated in Fig.~\ref{figelmod}(b), and
will not be discussed further.

If the electric modulation is strong enough, $w>2$, the function $\xi_0 (\xi) =
\xi -(w/2) \sin \xi$, Eq.~(\ref{pardarst1}), has extrema at $\xi_+ = \arccos
(2/w) >0$ and $\xi_- =- \xi_+$ (modulo $2\pi$) with values $\xi_0(\xi_{\pm}) =
\mp g(w/2)$, where $g(s)$ is defined by Eq.~(\ref{gvons}). Then, for $|\xi_0 |
\leq g(w/2)$ the equation $\xi_0(\xi)= \xi_0$ has three solutions $\xi$ in the
interval $|\xi| < \pi$, which are local extrema of the effective potential with
values 
\begin{equation}       \label{velextrem}
  V(\xi;\xi_0(\xi))/ V_{\rm cyc} = 1+(w/2)^2 - [1-(w/2)\cos \xi]^2 \, .
\end{equation} 
In order to find in the energy spectra the regions II corresponding to side
valleys of the effective potential, one may proceed as follows. One plots in the
extended zone scheme $V(\xi;\xi_0(\xi))$ versus $\xi_0(\xi)$, starting at
$\xi=-\pi$, where $\xi_0(\xi)= -\pi$ and $V(\xi;\xi_0(\xi))=-V_0$. With
increasing $\xi$, also $\xi_0(\xi)$ and $V(\xi;\xi_0(\xi))$ increase and reach
at $\xi= \xi_-$ their maximum values $g(w/2)$ and $V_{\rm cyc} (1+ w^2 /4)$,
respectively. As $\xi$ increases from $\xi= \xi_-$ to $\xi =0$, $\xi_0(\xi)$ and
$V(\xi;\xi_0(\xi))$ decrease towards the values $0$ and $V_0$, respectively.
Increasing $\xi$ from 0 to $\pi$ leads to the mirror image of the described
trace with respect to $\xi_0=0$: $\xi_0(\xi)= -\xi_0(-\xi)$ and
$V(\xi;\xi_0(\xi))=V(-\xi;\xi_0(-\xi))$. Finally these four line segments have
to be folded back into the ``first'' Brillouin zone $0 \leq \xi_0 \leq 2\pi$ to
obtain the absolute minimum of the effective potential as a function of $\xi_0$
and the boundaries of the regions II. In contrast to the strong magnetic
modulation, these regions become narrower with increasing energy and end at
$\xi_0 = \pm g(w/2)$ (modulo $2\pi$) with energy $E/ V_{\rm cyc} = 1+w^2 /4$.
For $g(w/2) > \pi$ the back-folding leads to an overlap of different branches of
the region II, that is the coexistence of back and forth running ``channeled''
states with ``drifting'' states in the same area of the $E$-$\xi_0$ diagram.
Figure~\ref{figelmod} shows for a typical example the quantum mechanical energy
spectrum together with the boundaries of region II obtained in this manner.  The
``very complicated'' energy spectrum obtained recently by Shi and
Szeto\cite{Shi96:12990} for strong electric modulation is thus explained by the
coexistence of channeled and drifting states.

In previous work \cite{Beton90:9229,Menne98:1707} it was pointed out that, for
given modulation period $a$ and strength $V_0$ and given energy $E=E_{\mathrm
  F}$, channeled orbits can exist only if the magnetic field $B_0$ is smaller
than a critical field $B_{\rm crit}$. Solving $E_{\mathrm F}/ V_{\rm cyc} =
1+w^2 /4$ for $E_{\mathrm F}>V_0 = w V_{\rm cyc}$ and $w >2$ with respect to the
magnetic field, one obtains the known result \cite{Menne98:1707}
\begin{equation}  \label{bcritic}
  B_{\rm crit} = \frac{2\pi V_0}{eav_{\mathrm F}} \,
  \left[\frac{2}{1+\sqrt{1-(V_0/E_{\mathrm F})^2}} \right]^{1/2} \, .
\end{equation}

\subsection{Weak magnetic modulation}
If a magnetic modulation is added to an electric one, very complicated
interference effects may result. Only if the phase shift $\varphi _s$ is zero or
$\pi$, the resulting energy spectrum will be symmetric in $\xi_0$. Even in that
case, the distribution of channeled states (regions II) in the $E$-$\xi_0$
diagram may become rather complicated, especially at low energies.  For the
mixed case channeled states may occur even if the modulation parameters $w$ and
$s$ are not large enough to produce them for the pure electric and the pure
magnetic modulation of these strengths. For weak magnetic modulation, $0< s <
1$, and arbitrary strength of the electric modulation, channeled states can
exist only below a certain energy, as in the pure electric modulation case.

For arbitrary phase shift $\varphi _s$ the energy spectrum may be so asymmetric
that in a certain energy range only channeled orbits exist which carry current
in one (say the positive $y$-) direction, but no channeled orbits carrying
current in the opposite direction. Such a situation is presented in
Fig.~\ref{figasymmetric}. The regions II, where channeled and drifting states
coexist, is again calculated from Eqs.~(\ref{pardarst1}) and (\ref{pardarst2}),
i.e.\ from purely classical arguments.

If a 2DEG is subjected to such an asymmetric mixed modulation, it may happen
that in the thermal equilibrium the channeled states carry a finite current. Of
course, this current must be compensated by a corresponding opposite current
carried by the drifting states.

For large magnetic modulation, $s>1$, and arbitrary electric modulation, the
magnetic modulation dominates the energy spectra at large energies. The regions
II with channeled states become more and more important, as can be seen from the
pole structure of Eqs.~(\ref{pardarst1}) and (\ref{pardarst2}). For very weak
electric modulation, the results reduce to those of the pure magnetic
modulation, apart from some peculiarities at very low energies, where additional
regions of channeled orbits may exist.

In magnetically modulated systems prepared by deposition of magnetic
micro-strips there is always an induced electric modulation due to the interface
stress between the ferromagnets and the semiconductor.  \cite {Ye95:3013} The
phase shift with respect to the magnetic modulation occurs when the external
magnetic field is tilted.\cite{Gerhardts96:11064} Nevertheless, in the known
experimental situations the stress potential amplitude is presumably much weaker
than our bare potential.

\section{Summary and discussion}
We have discussed in detail the quantum electronic states and energy spectra
$E_n(X_0)$ of a 2DEG in strong one-dimensional magnetic and electric
superlattices, and in a non-vanishing average external magnetic field. By
comparing the quantum results with the corresponding characteristics of the
classical motion, we achieved a detailed and intuitive understanding of the
energy spectra and eigenstates.  We found that the complicated parts of the
energy spectra (``regions II''), where branches with strong dispersion coexist
with those of low dispersion, coincide with the areas in the $E-X_0$ diagram in
which classically ``channeled'' orbits exist. 

For a systematic investigation of the possible energy spectra and eigen
states, and 
of the corresponding types of classical trajectories, it is useful to exploit
the scaling properties of the Hamiltonian. Then it is not necessary to vary
independently all the basic model paramaters, i.e.\ the strengths $B_m^0$ and
$V_0$ of magnetic and electric modulation, the modulation period $a$, and the
average magnetic field $B_0$. 
If one uses suitable units for energy and length, 
$V_{\rm cyc}$ and $a/ 2\pi$, respectively, one obtains the same classical
results and the same gross features of the energy spectra (the same position of
the regions~II), if one changes the {\em four} parameters  $B_m^0$, $V_0$,
$a$, and $B_0$ in such a manner that the {\em two} reduced modulation
strengths $s=B_m^0/B_0$ and $w=V_0/V_{\rm cyc}$ remain constant. For different
parameter sets with the same values of $s$ and $w$, only the density of the
energy bands is different in the plot of $E_n(X_0)/V_{\rm cyc} $ versus $K
X_0$, not its overall appearance. 
This is illustrated by Figs.~\ref{s_2_spec}(a) and \ref{s_5_spec}(a), for
which the regions II coincide. 
The reason for this behaviour is that, in these energy and
length units, the effective potential is invariant under the scaling
transformation $B_m^0 \rightarrow \gamma B_m^0$, $B_0  \rightarrow \gamma
B_0$, $a  \rightarrow \lambda a$ and $V_0 \rightarrow  \gamma^2 \lambda ^2
V_0$, for arbitrary positive $\gamma$ and $\lambda$. To leave the quantum
result exactly unchanged under a change of the four model parameters, one has
to keep $\alpha =(l_0 K)^4$ also unchanged. This is because only with the
restriction $\lambda = 1/ \sqrt{\gamma}$ the kinetic energy operator is also
independent of the scaling parameter $\gamma$ [see
Eq.~(\ref{schroedinger})]. Thus, in the suitable units, 
the exact quantum result depends only on {\em three} independent parameters
instead of {\em four}, and the characteristic classical features depend only
on {\em two}.

There is a close correspondence between the quantum states belonging to
strong-dispersion branches of the energy spectrum and the classical channeled
orbits. These orbits occur near lines of
vanishing total magnetic field or near minima of the electric modulation
potential and are restricted to individual side valleys of the effective
potential. They are always restricted to a part of a single
modulation period in $x$ direction and represent a fast motion along (wavy)
lines without self-intersections in positive or negative $y$ direction. The
corresponding quantum states are also essentially confined to the same space
region and belong to energy branches with strong dispersion. At a given value of
the constant of motion $X_0$, channeled orbits exist in energy intervals bounded
by adjacent relative minima and maxima of the effective potential, defining
bottom and top of the corresponding side valley. Plotting these classically
defined extrema versus $X_0$, one obtains the boundaries of the regions II of
the quantum energy spectrum.  Classically, for each channeled orbit there exists
a ``drifting'' orbit with the same constants of motion $X_0$ and $E$.  These
drifting orbits are self-intersecting trajectories which, for sufficiently large
energy, extend over more than one modulation period in $x$ direction and drift
slowly in $y$ direction. The corresponding quantum states belong to
low-dispersion branches of the energy spectrum. Quantum mechanically, the
channeled states do not appear at exactly the same energies as the drifting
states, and they usually have a larger energy spacing than the latter, since
they are confined to a narrower effective potential well.

We have demonstrated these features by model calculations based on simple
harmonic modulation fields. Qualitatively the obtained results and the methods
to derive them can easily be extended to more general modulation fields,
containing higher harmonics. This will be necessary, if the distance of the 2DEG
from the sample surface is not much larger than the period of the surface
structure creating the modulation.\cite{Gerhardts96:11064} Anharmonic effective
modulation potentials may also result from non-linear screening effects, even if
the bare modulation potential is
harmonic.\cite{Gossmann98:1680,Manolescu97:9707}

We have also performed several additional calculations and consistency checks
which are not documented in the main text.  E.g., we have checked the
equivalence of classical drift velocity and quantum group velocity beyond the
analytically accessible case of very weak modulation fields. For some examples
with strong modulation, we evaluated the quantum mechanical group velocity along
several energy bands $E_n(X_0)$ and compared the result with the drift velocity
of the corresponding classical trajectories with the same energy and $X_0$
values. For most parts of the bands the two velocities agreed perfectly.  A
systematic deviation was observed only in parameter regimes where the classical
trajectories are close to critical orbits, which have no quantum analog. Near
the critical orbits the modulus of the classical drift velocity increases rather
rapidly, whereas the quantum mechanical group velocity shows no anomaly.

We have also extended the band structure calculations to very strong magnetic
modulation ($B_m^0/B_0 =20$). While at high energies a complicated superpositon
of bands with steep and with flat dispersions, similar to that in
Fig.~\ref{s_6_spec}(a), was obtained, the bands at low energies tend to cluster
into groups separated by relatively large gaps. The low-energy part of the
spectrum was already reminiscent of the spectrum for vanishing average magnetic
field, where one obtains a one-dimensional Bloch energy spectrum for each value
of $p_y=-e B_0 X_0$.\cite{Ibrahim95:17321}

Concerning previous and forthcoming transport calculations, we conclude from the
close correspondence of the quantum and the classical approach, that at weak
average magnetic fields the classical calculations are appropriate, provided the
modulation fields are not too strong. For the very strong magnetic modulation
mentioned in the introduction, it may however happen, that the energy level
spacing of ``channeled orbits'' exceeds the thermal energy $k_B T$ in a regime
where $\hbar \omega_0 \ll k_B T$. Then we would expect modulation induced
quantum effects in the positive-magnetoresistance regime at low $ B_0$.

\acknowledgments We thank D.\ Pfannkuche for critical reading of the manuscript.
This work was supported by the German Bundes\-ministerium f\"ur Bildung und
Forschung (BMBF), Grant No. 01BM622.  One of us (A.M.) is grateful to the
Max-Planck-Institut f\"ur Festk\"orperforschung, Stuttgart, for support and
hospitality.


\begin{figure}
\caption{\label{illu1}
  (a): Effective potential $V(\xi;\xi_0)$ for magnetic cosine modulation with
  $s=B_m^0/B_0=0.5$ and $\xi_0/2\pi=1/4$. For a given energy $E_{\mathrm
    F}/V_{\mathrm cyc}=(KR_0)^2$ (horizontal lines) classical orbits exist where
  $V(\xi;\xi_0) \leq E_{\mathrm F}$. Solid line $E_{\mathrm F}=17.6 V_{\mathrm
    cyc}$, dotted line $E_{\mathrm F}=1.18 V_{\mathrm cyc}$.  (b): Locations of
  turning points $\xi_0^{\pm}(\xi)$ for the $E_{\mathrm F}$ values indicated in
  (a), same coding. Orbits with energy $E_{\mathrm F}$ and $\xi_0$ exist in an
  interval with $\xi_0^-(\xi) \leq \xi_0 \leq \xi_0^+(\xi)$.  (c): Corresponding
  orbits in $xy$--space, three cycles are shown each, the sense of motion is
  from filled to open dot.  }
\end{figure}
\begin{figure}
\caption{\label{illu2}
  As Fig.~\protect\ref{illu1}, but $\xi_0/2\pi=1/2$.  The effective potential is
  symmetric and therefore the guiding center of these drifting orbits does not
  drift in (c).  }
\end{figure}

\begin{figure}
  \caption{\label{figure1}(a) Landau bands for $s=0.5$. $B_0=0.2$ T and $a=800$
    nm, so that $V_{\mathrm cyc}=0.85$ meV and $\alpha=0.041$ and (b) total
    magnetic field $B(\xi_0)$.  The marked points on Landau bands 43 and 3 are
    the states for which the wave functions are shown in (c) and (d) in
    arbitrary units together with the corresponding effective potentials (dashed
    line). The wave functions are plotted with an offset, indicating the energy
    of the state.  The states of (c) and (d) are to be compared with the
    classical orbits in Figs.~\protect\ref{illu2} and \protect\ref{illu1}
    respectively.  }
\end{figure}
\begin{figure}
\caption{\label{figure2}(a) Energy spectrum for $s=1$. $B_0=0.1$ T and $a=800$
  nm, so that $V_{\mathrm cyc}=0.21$ meV and $\alpha=0.17$.  The dashed lines
  show $V(\xi_0,(2p+1)\pi)$ with $\mid p \mid \le 1$.  (b) Quantum mechanical
  (thick lines) and corresponding classical (thin lines) probability densities
  for two states. The chosen states are marked with dots in (a).  (c) Effective
  potentials and classical orbits for $E_{\mathrm F}=7V_{\mathrm cyc},
  \xi_0=0.5$ (solid lines) and $E_{\mathrm F}=39.6V_{\mathrm cyc}, \xi_0=3.18$
  (dotted lines).  The horizontal lines indicate the Fermi energy.  }
\end{figure}
\begin{figure}
\caption{(a): Effective potential $V(\xi;\xi_0)$ for magnetic cosine modulation
  with $s=B_m^0/B_0=2$, and $\xi_0/2\pi=0.2$. For given energy $E_{\mathrm F}$
  (horizontal lines) classical orbits exist where $V(\xi;\xi_0)\leq E_{\mathrm
    F}$.  (b): Total magnetic field.  (c): Locations of turning points
  $\xi_0^{\pm}(\xi)$ for the $E_{\mathrm F}$ values indicated in (a). The
  outermost pair of lines belongs to the largest, the innermost pair to the
  smallest $E_{\mathrm F}$ value. The constant of motion $\xi_0$ appears as a
  horizontal line in this plot ($\xi_0/2\pi=0.2$ is indicated). Orbits with
  fixed energy (i.~e. fixed curves $\xi_0^{\pm}(\xi)$) and this value of $\xi_0$
  exist in intervals with $\xi_0^-(\xi) \leq \xi_0 \leq \xi_0^+(\xi)$. Orbits,
  plotted in (d), with one turning point on $\xi_0^-(\xi)$ and the other on
  $\xi_0^+(\xi)$ are drifting orbits, the others are channeled orbits (see
  text).  }\label{illu3}
\end{figure}
\begin{figure}
\caption{As in Fig.~\protect\ref{illu3} but for $s=2$, and $\xi_0/2\pi=0.5$.
  Horizontal lines in (a) are for $E_{\mathrm F}/V_{\mathrm cyc}=40$ and
  $E_{\mathrm F}/V_{\mathrm cyc}=0.3$.  The inset shows $V(\xi;\xi_0)$ enlarged
  between $\xi=0$ and $\xi=2\pi$, where it has three zeroes. (c): For both
  indicated energies there exist three orbits, one drifting and two channeled
  orbits for $E_{\mathrm F}/V_{\mathrm cyc}=40$, and three drifting orbits for
  $E_{\mathrm F}/V_{\mathrm cyc}=0.3$, plotted in (d).  Since the effective
  potential is symmetric, there is no guiding center drift for the central
  drifting orbits.  }\label{illu4}
\end{figure}
\begin{figure}
\caption{\label{s_2_spec}(a) Energy spectrum (first 75 bands) for $s=2$.
  $B_0=0.05$ T and $a=800$ nm, so that $V_{\mathrm cyc}=0.053$ meV and
  $\alpha=0.67$.  Effective potential and specific states (b) ($n$=0,20,22) for
  $\xi_0/2\pi=0.5$ and (c) ($n$=0,1,2,25,44,45) for $\xi_0/2\pi=0.2$.  }
\end{figure}
\begin{figure}
  \caption{\label{s_5_spec}(a) Energy spectrum (first 100 bands) for $s=2$. 
    $B_0=0.2$ T and $a=800$ nm, so that $V_{\mathrm cyc}=0.851$ meV and
    $\alpha=0.041$.  (c) Effective potential and specific states
    ($n$=0,2,3,4,5), marked in extract of spectrum (b), for $\xi_0/2\pi=0.5$.  }
\end{figure}
\begin{figure}
\caption{\label{s_6_spec}(a) Energy spectrum for $s=5$, $B_0=0.1$ T and $a=800$
  nm, so that $V_{\mathrm cyc}=0.213$ meV and $\alpha=0.16$.  (b) Total magnetic
  field.  (c) Effective potential and specific states for $\xi_0/2\pi=0.016$.
  Typical channeled states $n=16$ and $n=22$ and a narrow drifting states,
  $n=19$.  (d) Effective potential and specific states for $\xi_0/2\pi=0.493$.
  Wide drifting state, $n=17$, narrow drifting states, $n=5$ and $n=6$, and a
  channeled state, $n=7$.  }
\end{figure}
\begin{figure} 
\caption{\label{figelmod}Pure electric modulation of strength $w=20$. 
  (a) Energy spectrum (first 80 bands) for $B_0=0.05$ T and $a=800$ nm, so that
  $V_{\mathrm cyc}=0.053$ meV and $\alpha=0.67$.  (b) Drifting orbit (solid
  line) and channeled orbit (dotted line) for $\xi_0/2\pi=0.3$.  The energies
  (see marked states in (a)) are chosen in such a way, that $\xi_c<\xi_0$ for
  the drifting and $\xi_c>\xi_0$ for the channeled orbit, see text.  }
\end{figure}
\begin{figure} 
\caption{\label{figasymmetric}
  (a) The first 175 Landau bands for combined magnetic and electric modulations.
  $B_0=0.1$ T and $a=800$ nm, so that $V_{\mathrm cyc}=0.213$ and $\alpha=0.16$.
  The modulation strengths are $s=0.8$ and $w=14.3$, the relative phaseshift is
  $\pi/2$.  (b) $\xi_0^{\pm}$-curves for $E/V_{\mathrm cyc}=110$. At this energy
  $\xi_0^{+}$ has local extrema, while there are none in $\xi_0^{-}$. Therefore
  all channeled orbits have negative velocities $v_y$.  }
\end{figure}
\onecolumn
\newpage
\Huge Fig.\ref{illu1} \normalsize Zwerschke\\
\includegraphics{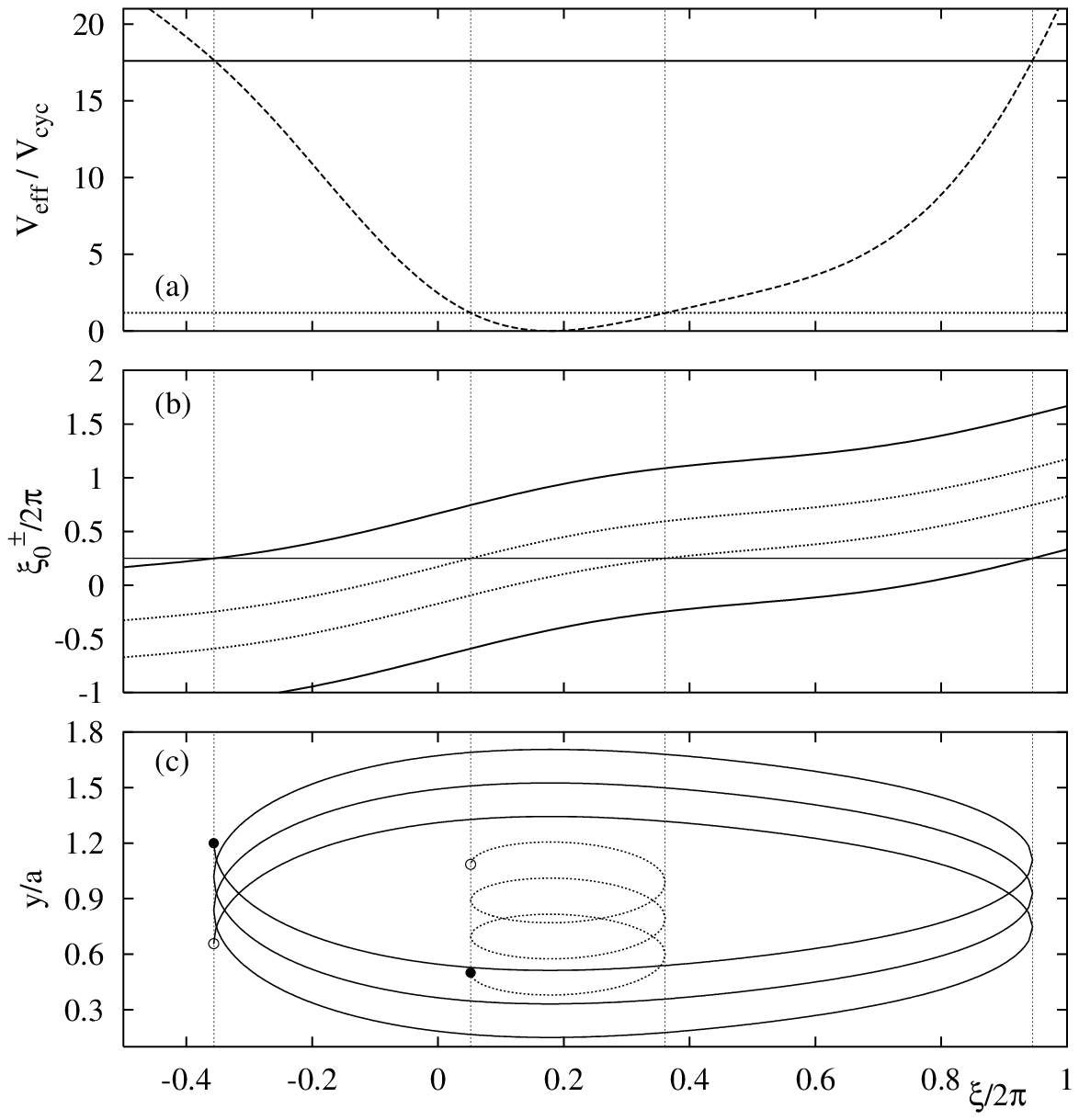}
\vfill
\newpage
\Huge Fig.\ref{illu2} \normalsize Zwerschke\\
\includegraphics{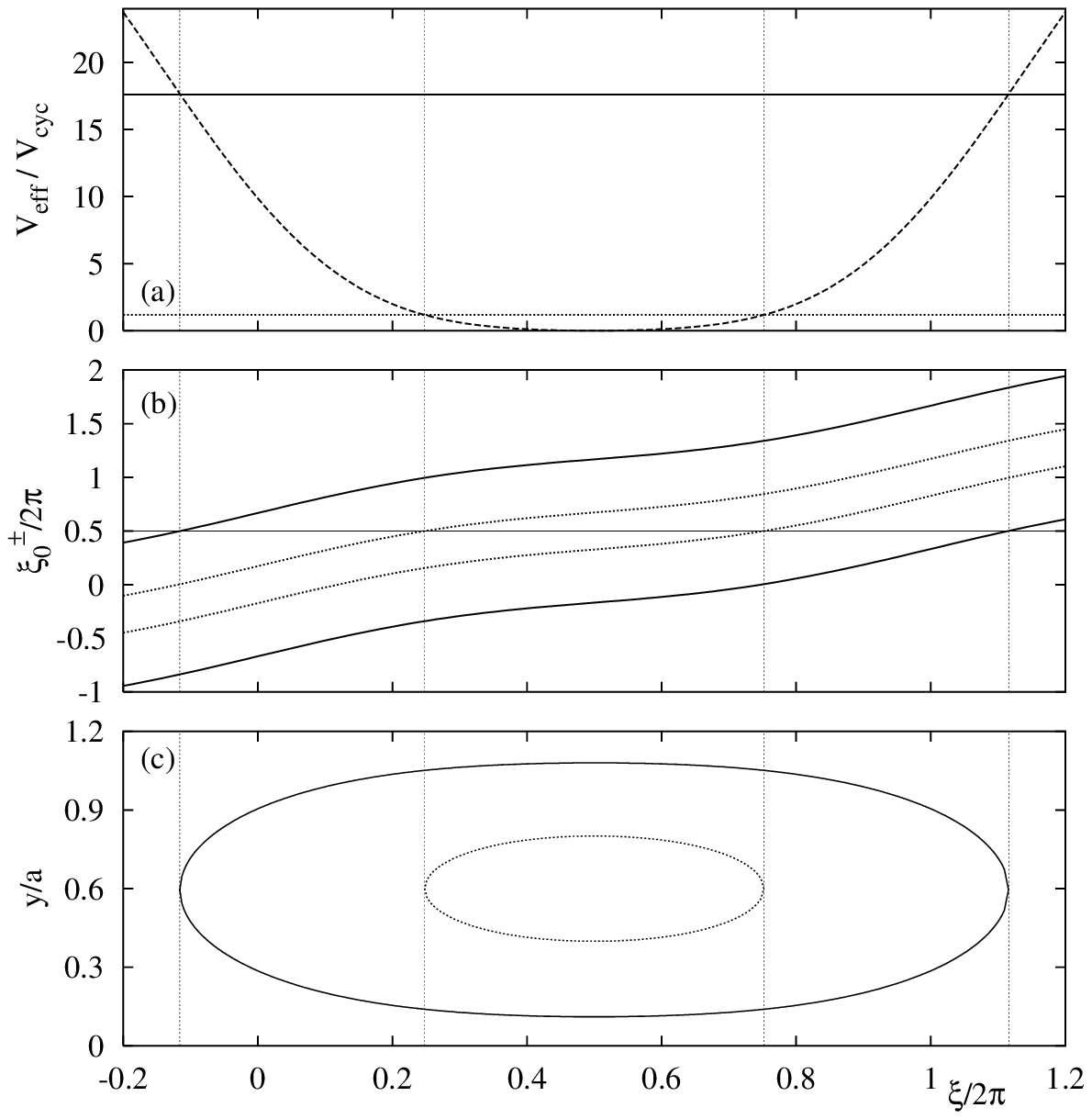}
\vfill
\newpage
\Huge Fig.\ref{figure1} \normalsize Zwerschke\\
\includegraphics{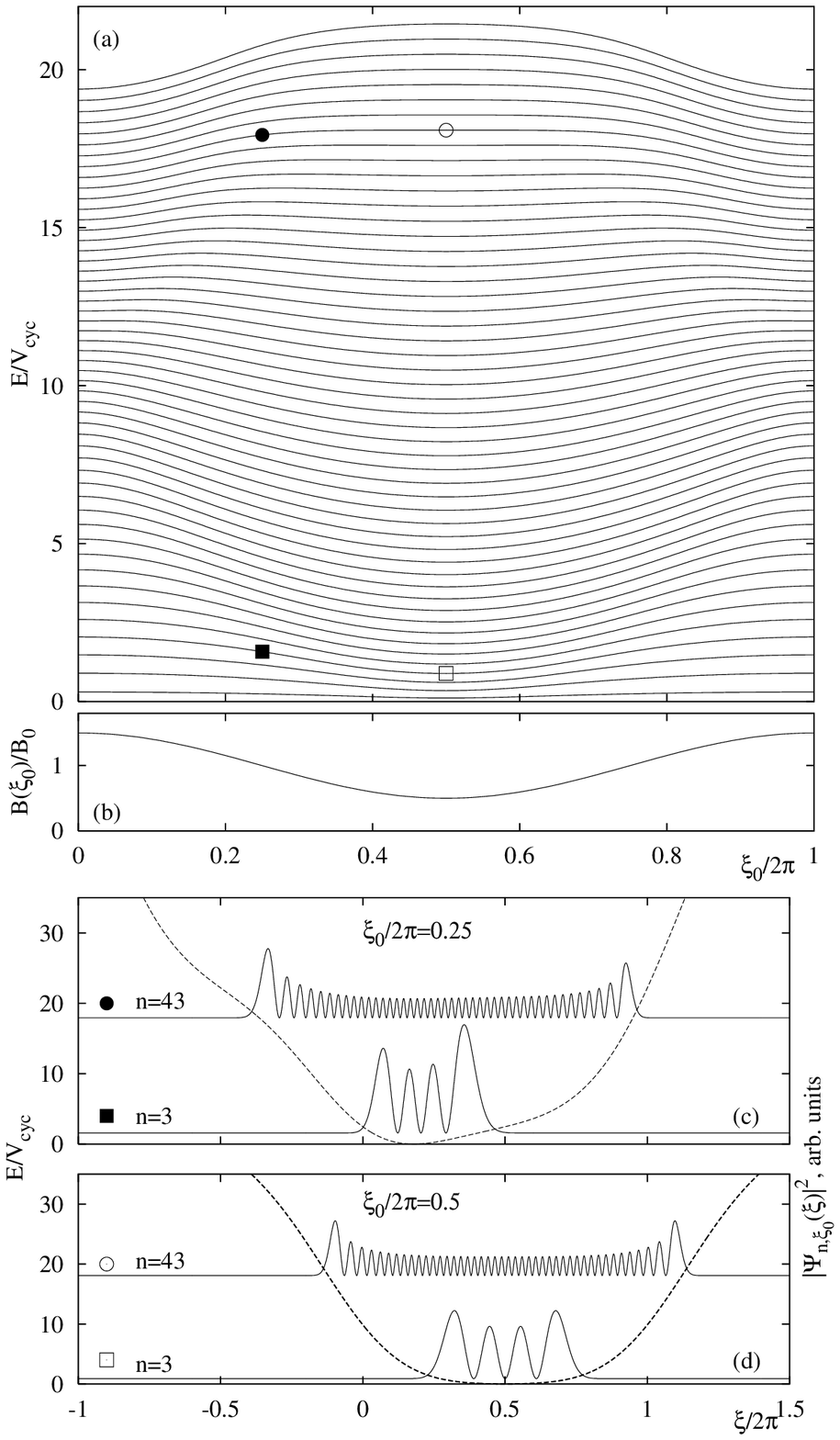}
\vfill
\newpage
\Huge Fig.\ref{figure2} \normalsize Zwerschke\\
\includegraphics{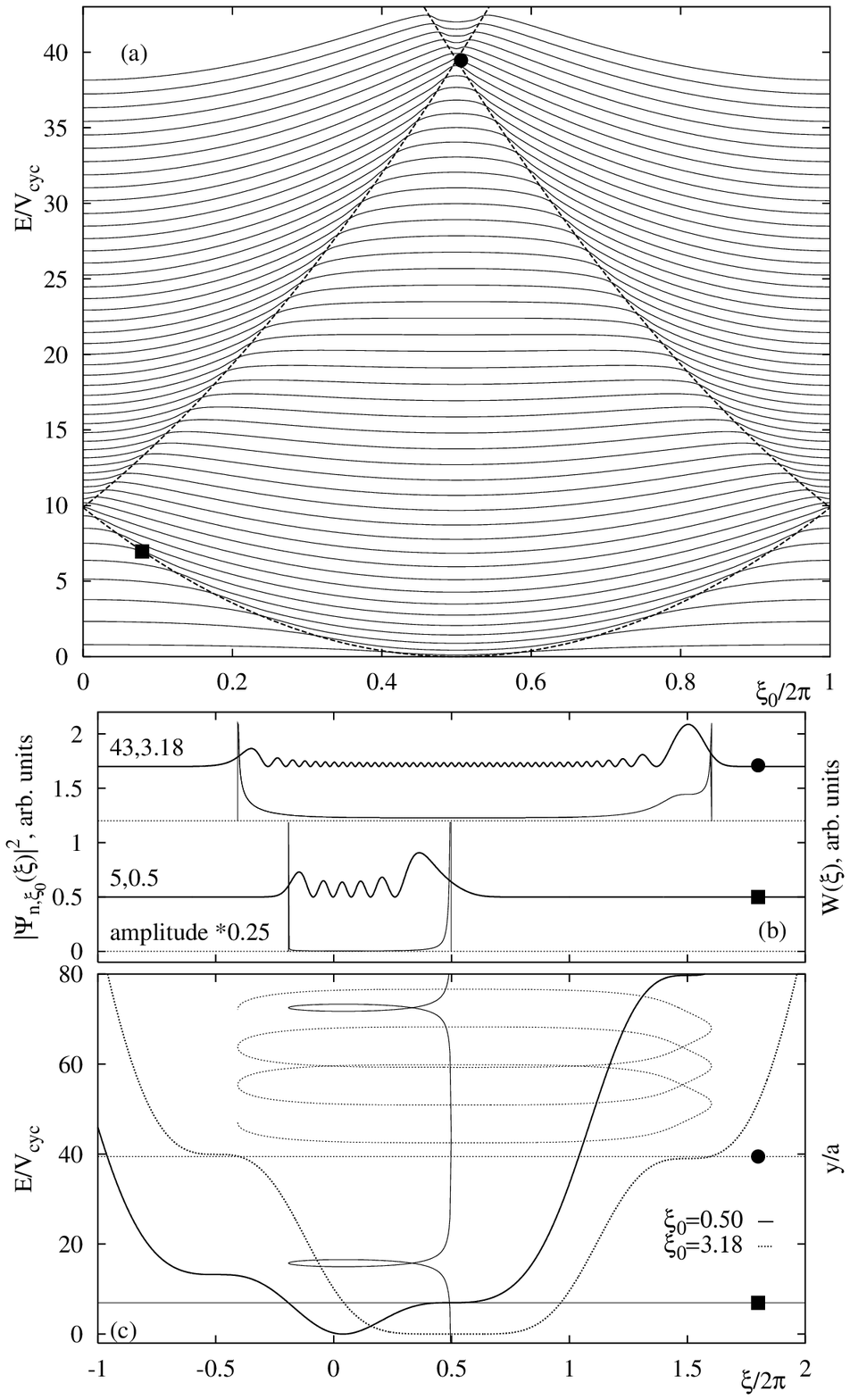}
\vfill
\newpage
\Huge Fig.\ref{illu3} \normalsize Zwerschke\\
\includegraphics{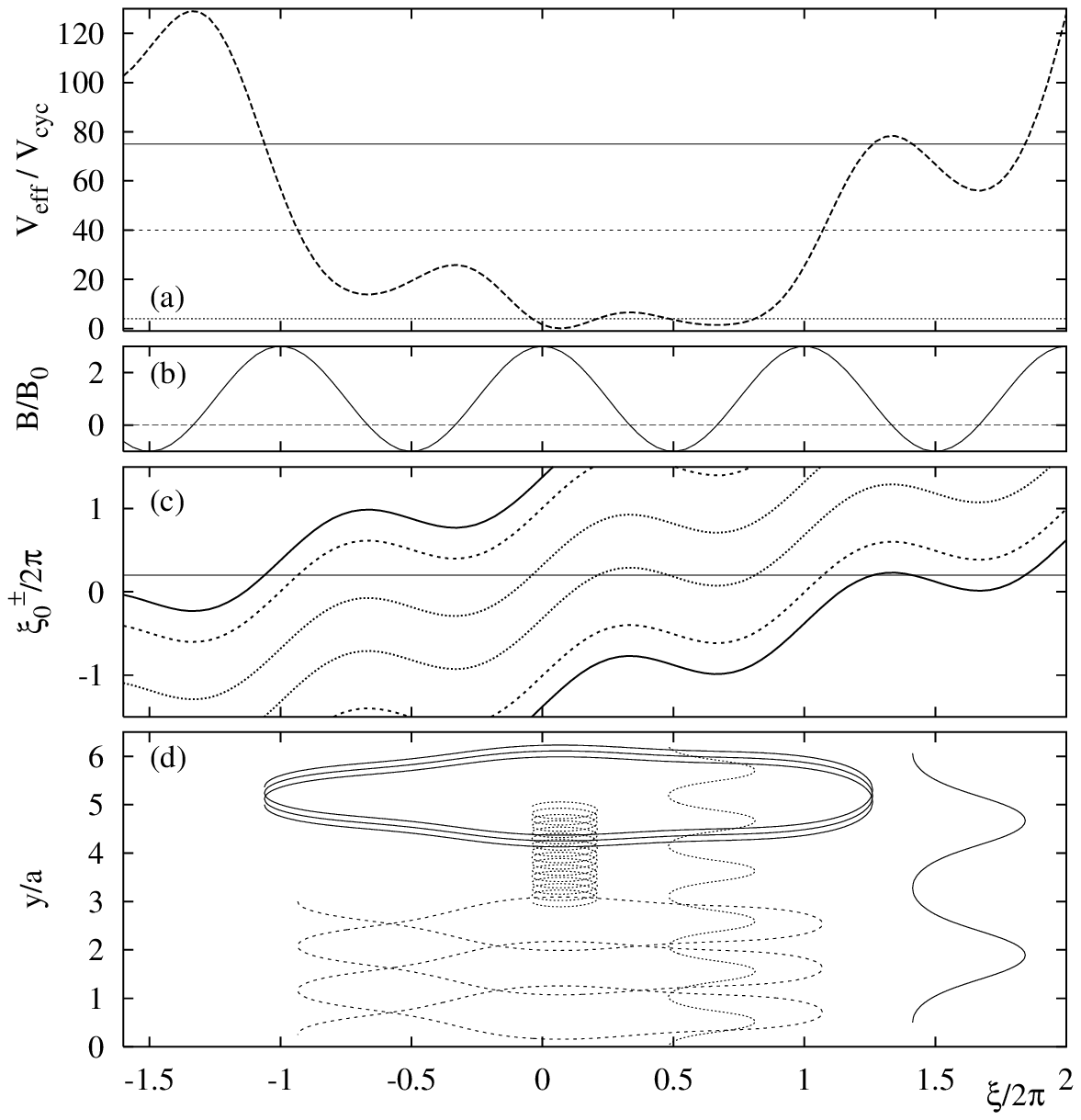}
\vfill
\newpage
\Huge Fig.\ref{illu4} \normalsize Zwerschke\\
\includegraphics{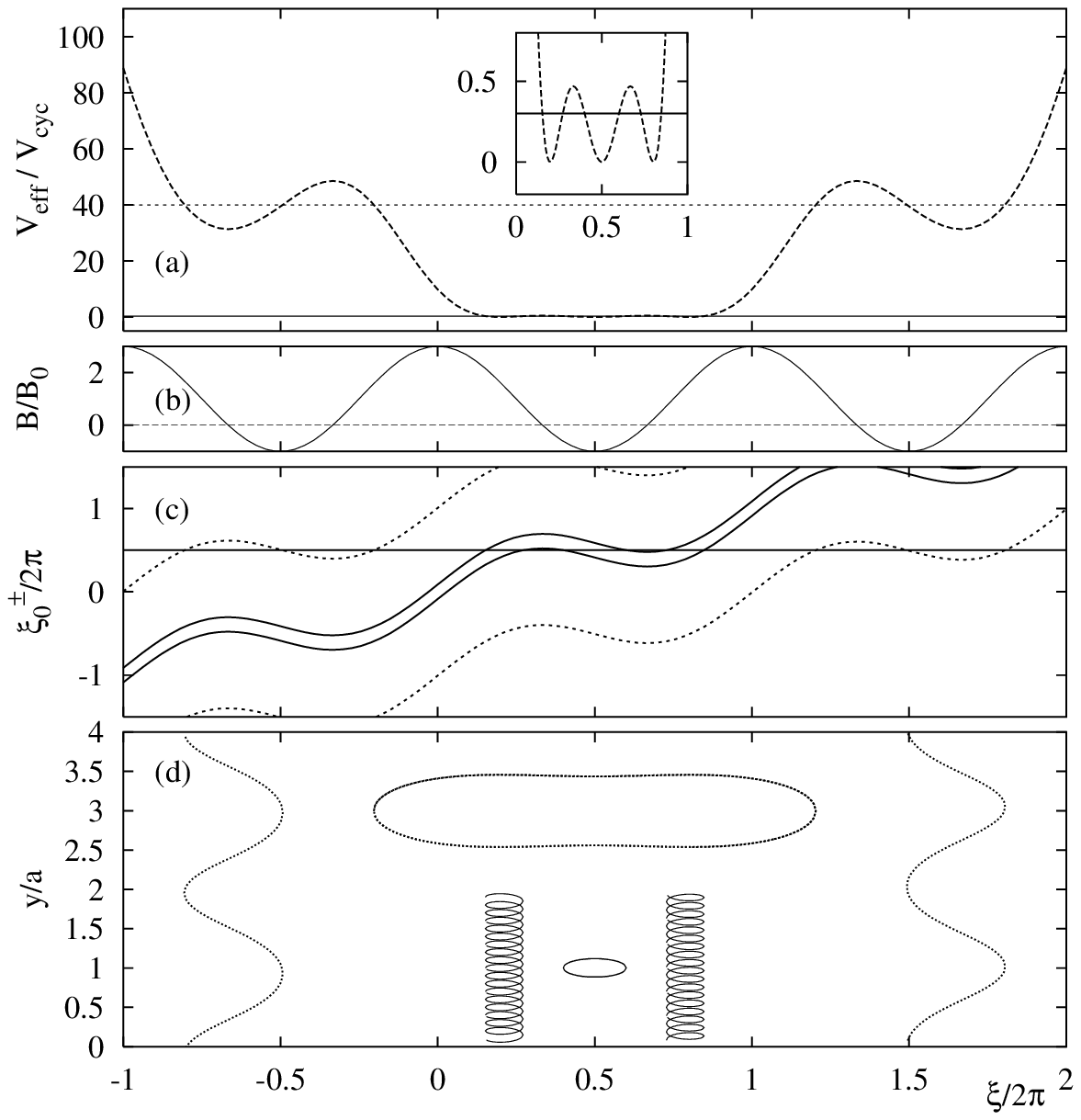}
\vfill
\newpage
\Huge Fig.\ref{s_2_spec} \normalsize Zwerschke\\
\includegraphics{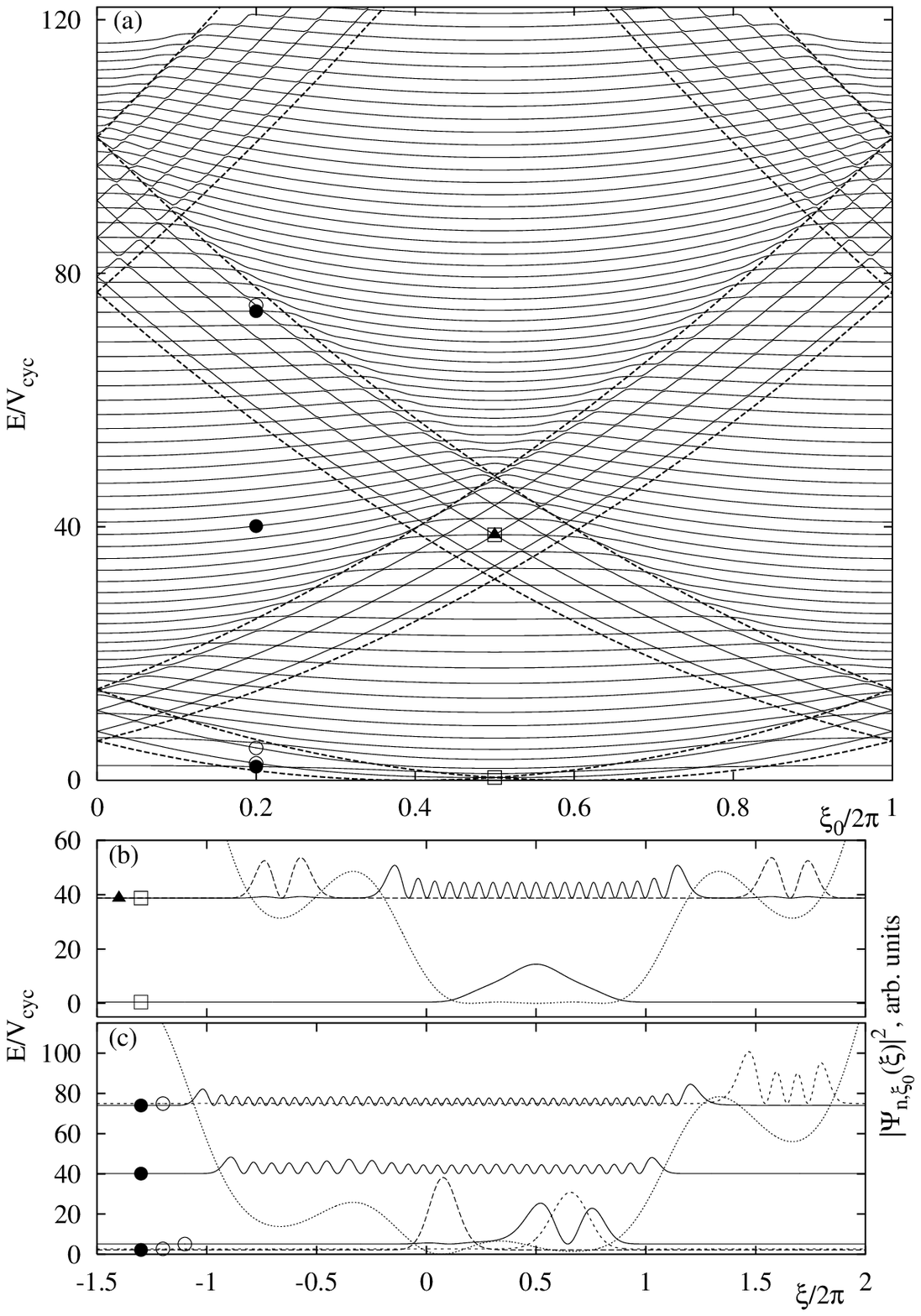}
\vfill
\newpage
\Huge Fig.\ref{s_5_spec} \normalsize Zwerschke\\
\includegraphics{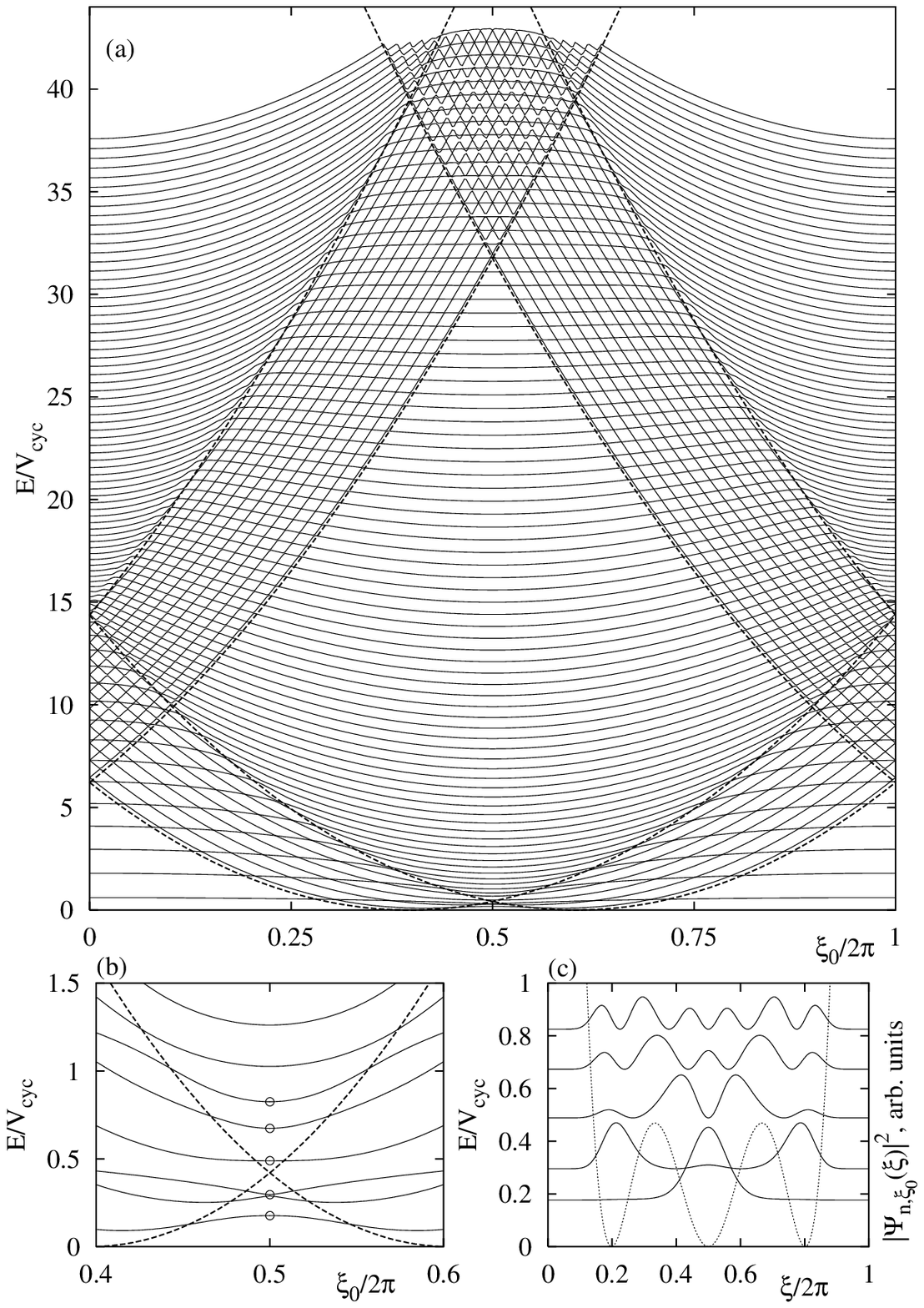}
\vfill
\newpage
\Huge Fig.\ref{s_6_spec} \normalsize Zwerschke\\
\includegraphics{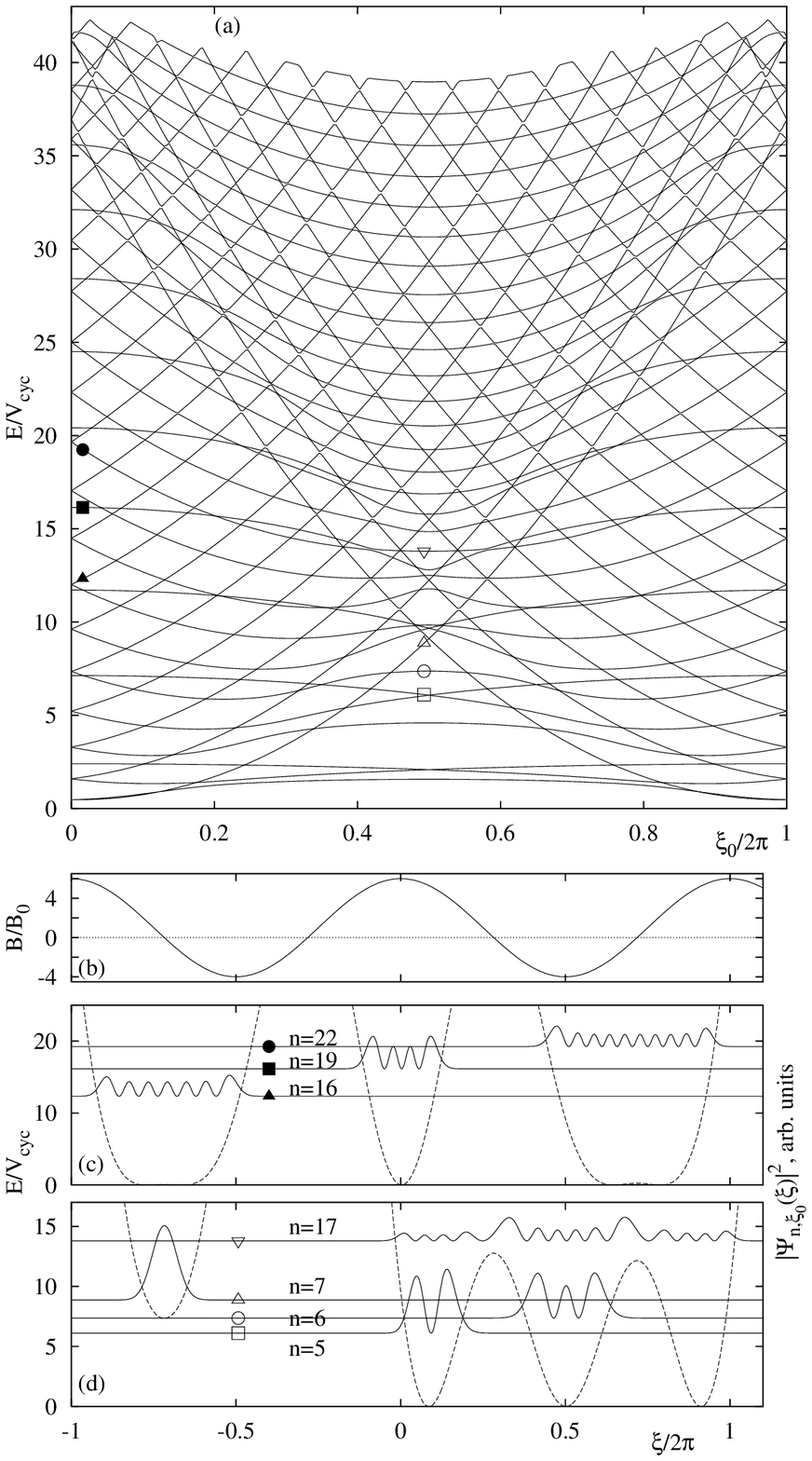}
\vfill
\newpage
\Huge Fig.\ref{figelmod}  \normalsize Zwerschke\\
\includegraphics{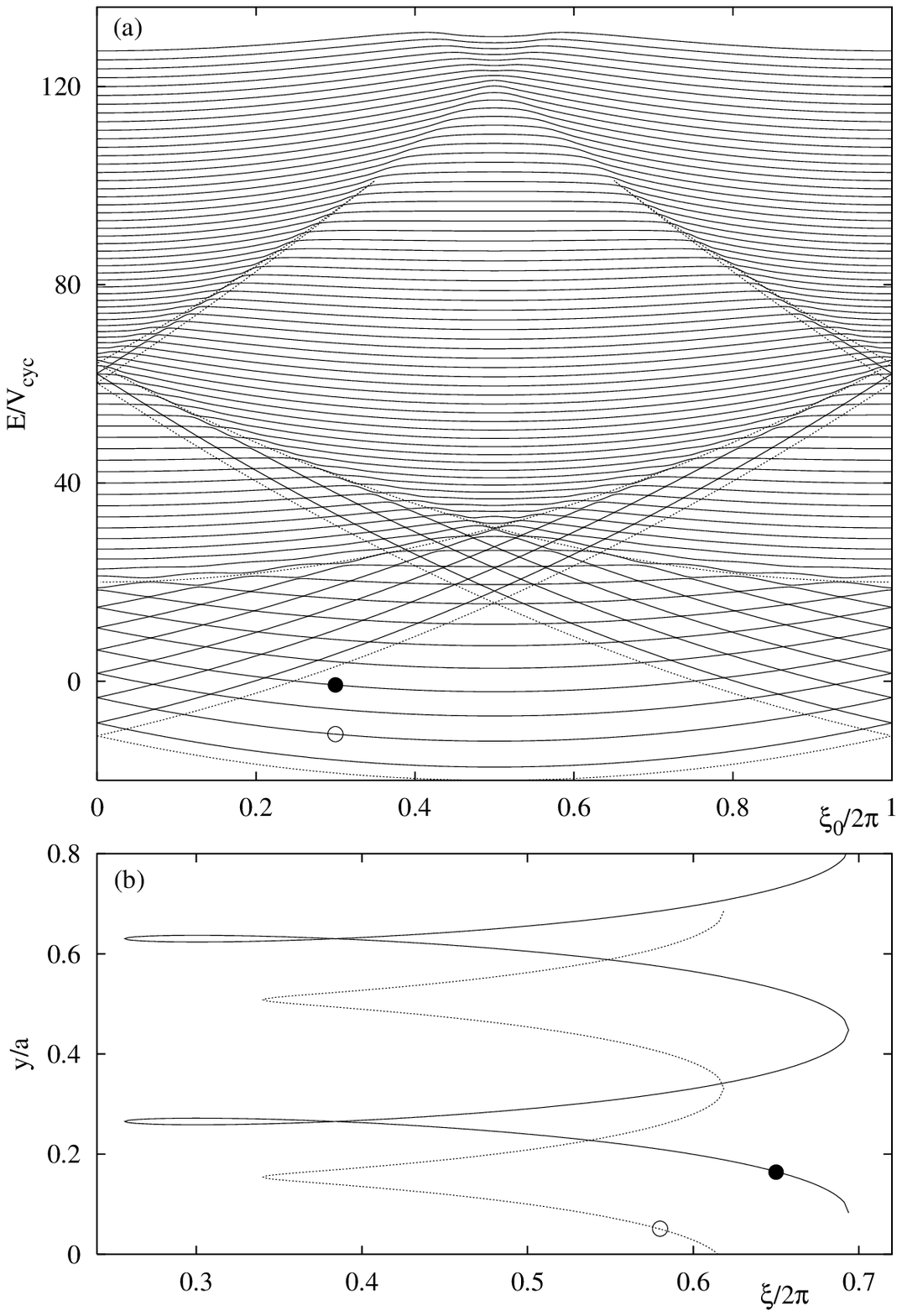}
\vfill
\newpage
\Huge Fig.\ref{figasymmetric} \normalsize Zwerschke\\
\includegraphics{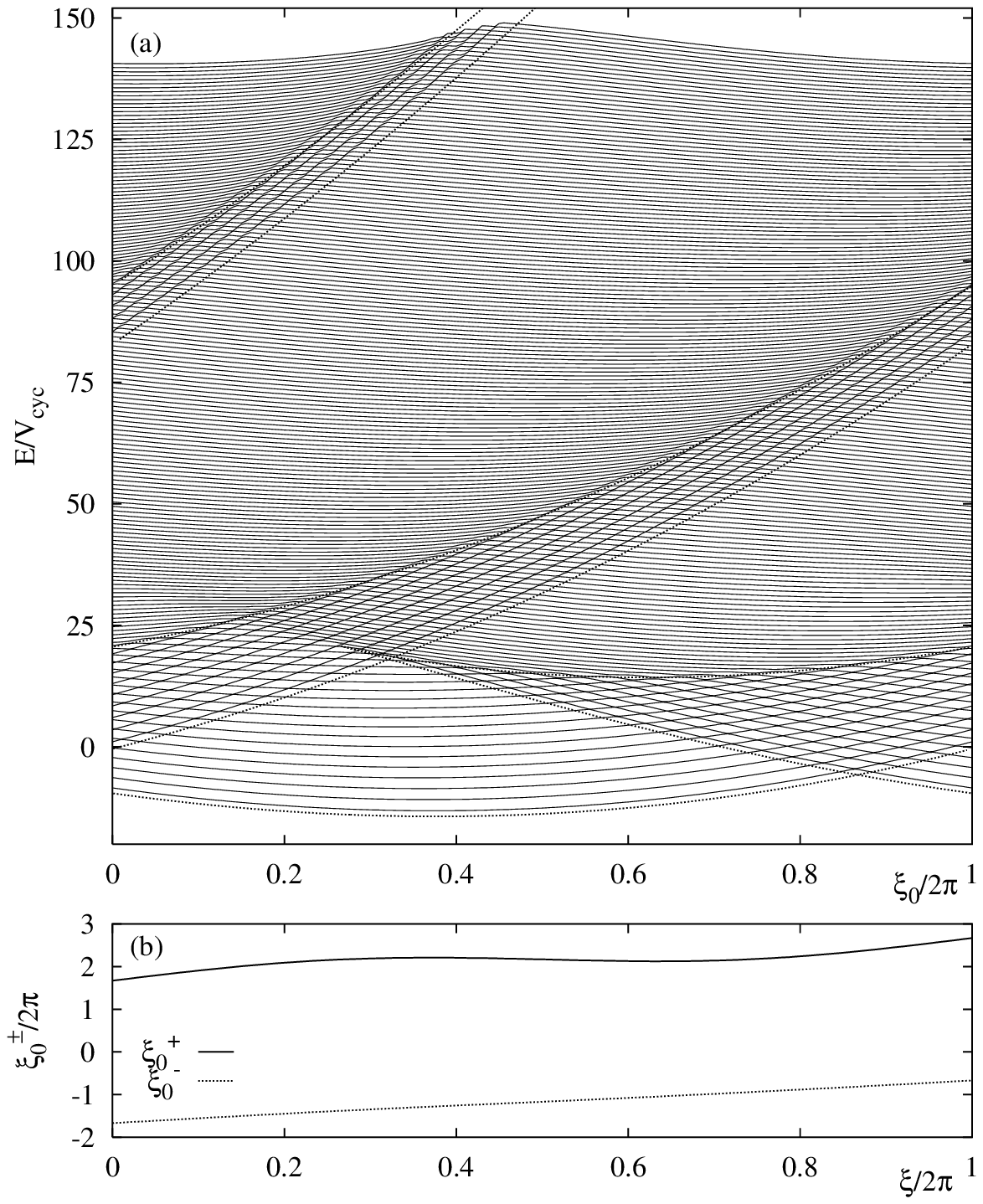}
\vfill
\end{document}